\documentclass[conference]{IEEEtran}
\IEEEoverridecommandlockouts
\usepackage{amsmath,amssymb,amsfonts}

\usepackage{graphicx}
\usepackage{textcomp}
\usepackage{xcolor}
\usepackage{subfig}
\usepackage{textcomp}
\usepackage{xcolor}
\usepackage{tikz}
\usepackage{quantikz}
\usepackage{algorithm}
\usepackage{algorithmicx}
\usepackage{float}

\usepackage{algpseudocode}%

\usepackage{pgffor}
\usepackage{soul}
\usepackage[sorting=none, doi = false, isbn = false, style = ieee]{biblatex} % for a biblio
\usepackage{csquotes}
\usepackage{hyperref}% add hypertext capabilities
\usepackage{float}
\usepackage[english]{babel}
\usepackage{amsthm}
\usepackage{amssymb}
\usepackage{enumitem}
\usepackage{dsfont}
\usepackage{tikz}
\usepackage{multirow}%
\usepackage{subfig}
\usepackage{bm}
\usepackage{bbold}

\usetikzlibrary{shapes.geometric, arrows, quantikz2}

\def\BibTeX{{\rm B\kern-.05em{\sc i\kern-.025em b}\kern-.08em
    T\kern-.1667em\lower.7ex\hbox{E}\kern-.125emX}}

\begin{document}

\title{Scheduling Quantum Annealing for Active User Detection in a NOMA Network}

\author{\IEEEauthorblockN{Romain Piron}
\IEEEauthorblockA{\textit{Universit\'e de Lyon, INSA Lyon, INRIA} \\
\textit{CITI EA 3720}\\
69621 Villeurbanne, France \\
romain.piron@insa-lyon.fr}
\and
\IEEEauthorblockN{Claire Goursaud}
\IEEEauthorblockA{\textit{Universit\'e de Lyon, INSA Lyon, INRIA} \\
\textit{CITI EA 3720}\\
69621 Villeurbanne, France \\
claire.goursaud@insa-lyon.fr}

}

\maketitle

\begin{abstract}
Active user detection in a non-orthogonal multiple access (NOMA) network is a major challenge for 5G/6G applications. However, classical algorithms that can perform this task suffer either from complexity or reduced performances. This work aims at proposing a quantum annealing approach to overcome this trade-off. Firstly, we show that the \textit{maximum a posteriori} decoder of the activity pattern of the network can be seen as the ground state of an Ising Hamiltonian. For N users in a network with perfect channels, we propose a universal control function to schedule the annealing process. Our approach avoids to continuously compute the optimal control function but still ensures high success probability while demanding a lower annealing time than a linear control function. This advantage holds even in the presence of imperfections in the network.
\end{abstract}


\section{Introduction}

One of the major challenges of 5G wireless networks is to achieve ultra-reliable low latency communications (uRLLC) requirements to serve applications as the Industrial Internet of Things \cite{popovski_wireless_2019}. However, the allocation of a dedicated time slot or frequency prior to transmission for each user is no longer adapted for typical uRLLC networks because of their growing size \cite{akpakwu_survey_2018,lee_6g_2021}. Instead, each time that the users wish to transmit a signal, they must complete a random access procedure \cite{chetot_active_2023}. 

This procedure begins by an activity notification sent by the active users in the uplink channel. To do so, the code division multiple access (CDMA) approach can be used \cite{fletcher_-off_2009}. An identification sequence is attributed to each node of the network and all active users simply transmit their code. This approach permits sharing a common time-frequency resource to all the users without interference.

It recently appeared that leveraging non orthogonal multiple access (NOMA) in CDMA systems is promising \cite{dai_survey_2018}. In practice, it allows to  attribute non orthogonal identification sequences to the nodes. Thus, one can reduce the size of these sequences for a given size of the users' set.

Detecting the active users $-$ or the activity pattern $-$ accurately from the received signal is called active user detection (AUD) \cite{chetot_active_2023}. It poses a significant challenge in NOMA systems since the cross-correlations between the sequences do not vanish. Thus, the users cannot be detected separately. While methods like the zero-forcing technique \cite{xiao_quantifying_2010} or traditional correlation receiver decoders \cite{goursaud_ds-ocdma_2006} offer computational efficiency, their performance is limited. Conversely, the maximum a posteriori (MAP) Bayesian estimator \cite{zdeborova_statistical_2016} is the most reliable but computationally demanding, being NP-hard to compute.

In AUD scenarios in which the channel state is known, one can show that the computation of the MAP estimator is a quadratic unconstrained binary optimization (QUBO) \cite{hauke_perspectives_2020} problem. The equivalence between QUBO problems and Ising Hamiltonians allows one to formulate the AUD problem as a ground state searching problem. However, an exhaustive search over the space of configurations to find the ground state has an exponential cost with the system size. 

Several methods have been investigated to avoid this exhaustive search.  Quantum annealing (QA) originally introduced by Kadowaki and Nishimori as a quantum analogue of thermal annealing in \cite{kadowaki_quantum_1998} is considered today as a promising quantum computing-based strategy for the ground state searching of several classes of Hamiltonians \cite{hauke_perspectives_2020, seki_quantum_2012}. Like many authors today, we use the term quantum annealing to refer to QA-inspired adiabatic quantum computation \cite{hauke_perspectives_2020, brady_behavior_2021}. Much work remains to be done regarding execution time, but QA already outperformed classical thermal annealing in reaching the ground state of some classes of Hamiltonians \cite{hauke_perspectives_2020}. 

The control function that introduces transverse quantum fluctuations in QA usually needs to be adapted according to the parameters of the target Hamiltonian \cite{roland_quantum_2002}. Unfortunately, in typical AUD models several variables are not deterministic but rather random variables \cite{chetot_active_2023}. This would require to compute the control function for each received signal. This is why, in this paper, we propose a strategy to build a generic control function to parameterize a QA process for AUD given the size $N$ of the wireless network.

The rest of the paper is organized as follows. Sec. \ref{sec:QAforAUD} defines our model and build the Ising Hamiltonian associated with the MAP estimator for AUD. In Sec. \ref{sec:reviewQA}, we briefly review the adiabatic theorem and show how to derive a control function for QA that ensures adiabaticity locally during the annealing process. Then, in Sec. \ref{sec:controlFctNoChannel} we adopt a simplified scenario of our model to propose a universal control function to schedule a QA approach for AUD once the size of the network has been fixed. Finally, Sec. \ref{sec:controlFctChannel} takes into account the channel state information in our AUD model. 

\subsection*{Notations}

The set of matrices over a field $\mathbb{K}$ of size $N \times M$ is denoted $\mathbb{K}^{N \times M}$ with the usual identification $\mathbb{K}^{N \times 1} \equiv \mathbb{K}^N$. We respectively denote a scalar, vector, matrix by $x$, $\bm{x}$, $\bm{X}$. 

The diagonal matrix whose diagonal elements correspond to the components of a vector $\bm{x}$ is denoted $\text{diag}(\bm{x})$. We also introduce $\bm{1}_N = [1, \dots, 1] \in \mathbb{K}^N $, finally the identity matrix of size $M\times M$ corresponds to $I_M = \text{diag}(\bm{1}_M)$.

The multivariate normal distribution with the mean vector $\bm{\mu}$ and the covariance matrix $\bm{\Sigma}$ is denoted $\mathcal{N}(\bm{\mu},\bm{\Sigma})$. The Dirac distribution picked at $x$ is denoted $\text{Dirac}(x)$. The uniform distribution over a set $\Omega$ is denoted $\mathcal{U}(\Omega)$.

\section{QA formulation for AUD}
\label{sec:QAforAUD}
\subsection{Context and problem defintion}

We consider a network with $N$ users. Each user $\#i$ is equipped with a single antenna and assigned with a bipolar identification sequence \cite{karafolas_optical_1995} $-$ also called code $-$ of length $M$ denoted $\bm{c_i}$. This code constitutes a signal that is transmitted by the $i$-th user. Its power is given by:
\begin{align}
    P_{\text{signal},i} = \lVert \bm{c}_i \rVert^2.
\end{align}
It is often convenient to work with normalized signals such that $P_{\text{signal},i} = 1$. Hence, the bipolar codes are chosen in the set:
\begin{align}
    \bm{c}_i \in \left\{-\frac{1}{\sqrt{M}}, \frac{1}{\sqrt{M}}\right\}^M.
\end{align}
These vectors must also satisfy a quasi-linearly independence to ensure the decodability, which is discussed in Appendix \ref{seqChoice}.

Let us assume that a random subset $\mathcal{N}_A$ of these users wish to transmit their code to a base station (BS) to notify their activity on the wireless channel. This subset constitutes the activity pattern of the network encoded in the bit-string $\bm{b}^{(0)} \in \{0,1\}^N$ so that one can write:
\begin{align}
    \mathcal{N}_A = \left\{i \in \{1, \dots, N\} \,\, \, \text{s.t} \,\,\, b_i^{(0)} = 1\right\}.
\end{align}

Each propagation path between the $i$-th user and the BS might suffer from attenuation due to the Rayleigh fading coming from the isotropic scattering assumption \cite{marzetta_fundamentals_2016}. It is modelled by $N$ fading coefficients $(w_i)_{i=1,\dots,N}$ that obey a different probability law depending on the quality of the channels:
\begin{align}
        \bm{w} \sim \begin{dcases}
            \text{Dirac}(\bm{1}_N) & \,\, \text{for perfect channels}\\
            \mathcal{N}(\bm{0},I_N) & \,\,\text{for imperfect channels}
        \end{dcases}
    \label{lawForW}
\end{align}
where $\bm{w} = (w_i)_{i=1,\dots,N}$. At the end, one measures the following signal at the receiver side:
\begin{align}
    \bm{y} = \sum_{i=1}^N w_i b_i^{(0)} \bm{c}_i + \bm{n},
    \label{receivedY}
\end{align}
where $\bm{n} \sim \mathcal{N}(\bm{0},\xi^2 I_M)$ is an additive white Gaussian noise that models the thermal noise at the BS.

The AUD problem consists in retrieving the activity pattern $\bm{b}^{(0)}$ of the network with the knowledge of the received signal $\bm{y}$ and the fading coefficients $\bm{w}$. As mentioned in the introduction, the MAP estimator is the most reliable decoder for such task. It is defined by:
\begin{align}
    \hat{\bm{b}}^{\text{MAP}} = \underset{\bm{b} \in \{0,1\}N}{\text{arg max }}     \mathbb{P}_{\bm{\mathsf{b}}|\bm{\mathsf{y}},\bm{\mathsf{w}}}(\bm{b}|\bm{y},\bm{w})
\end{align}
In the next part, we express the probability mass function $\mathbb{P}_{\bm{\mathsf{b}}|\bm{\mathsf{y}},\bm{\mathsf{w}}}(\bm{b}|\bm{y},\bm{w})$ in terms of the variables $b_i$.

\subsection{QUBO formulation for the MAP detector}

Eq. \ref{receivedY} implies that $f_{\bm{\mathsf{y}}|\bm{\mathsf{b}},\bm{\mathsf{w}}}$ simply corresponds to the probability density function of a Gaussian law. In addition, the Bayes' formula states that \cite{chetot_activity_nodate}:
\begin{align}
    \mathbb{P}_{\bm{\mathsf{b}}|\bm{\mathsf{y}},\bm{\mathsf{w}}}(\bm{b}|\bm{y},\bm{w}) f_{\bm{\mathsf{y}}|\bm{\mathsf{w}}}(\bm{y}|\bm{w}) = f_{\bm{\mathsf{y}}|\bm{\mathsf{b}},\bm{\mathsf{w}}}(\bm{y}|\bm{b},\bm{w})\mathbb{P}_{\bm{\mathsf{b}}}(\bm{b}).
\end{align}
From this point, a simple hypothesis consists in assuming a uniform distribution over the activity patterns. In this scenario, the prior distribution $\mathbb{P}_{\bm{b}}$ is constant which yields:
\begin{align}
    \mathbb{P}_{\bm{\mathsf{b}}|\bm{\mathsf{y}},\bm{\mathsf{w}}}(\bm{b}|\bm{y},\bm{w}) \propto \exp\left(- \frac{\left\lVert\bm{y}-\bm{C}\cdot \text{diag}(\bm{w}) \cdot \bm{b}\right\rVert^2}{2 \xi^2}\right),
    \label{successProbaMAP}
\end{align}
where the sequences have been concatenated in the matrix $\bm{C} = [\bm{c}_1 , \dots, \bm{c}_N] \in \mathbb{C}^{M \times N}$. Hence the MAP detector reads:
\begin{align}
    \hat{\bm{b}}^{\text{MAP}} = \underset{\bm{b} \in \{0,1\}^N}{\text{arg min }} \left\{\left\lVert\bm{y}-\bm{C}\cdot \text{diag}(\bm{w}) \cdot \bm{b}\right\rVert^2\right\}.
\end{align}
The computation of $\hat{\bm{b}}^{\text{MAP}}$ aims at recovering $\bm{b}^{(0)}$ with $(\bm{y},\bm{w})$ as inputs. Developing this objective function yields to the expression:
\begin{align}
    \begin{split}
        \left\lVert\bm{y}-\bm{C}\cdot \right.&\left. \text{diag}(\bm{w}) \cdot \bm{b}\right\rVert^2 = \lVert \bm{y} \rVert^2 - 2 \sum_{i=1}^N (\bm{y}\cdot w_i \bm{c}_i) b_i \\
        & + \sum_{i,j=1}^N (w_i \bm{c}_i \cdot w_j \bm{c}_j )b_i b_j.
    \end{split}
    \label{QUBOformML}
\end{align}
Once one throws away the irrelevant constant $\lVert\bm{y}\rVert^2$, our minimization problem corresponds to the general form
\begin{align}
    \hat{\bm{b}}^{\text{MAP}} = \underset{\bm{b} \in \{0,1\}^N}{\text{arg min }} \left\{ \sum_{i,j} Q_{ij} b_i b_j + \sum_{i} a_i b_i \right\}.
\end{align}
Thus, the MAP detector can be written as a QUBO problem \cite{hauke_perspectives_2020} expressed in terms of the binary variables $b_i$. 

\subsection{Building the Ising Hamiltonian}

As underlined by the authors of \cite{hauke_perspectives_2020}, a QUBO problem can be mapped to an Ising Hamiltonian by the change of variable $\sigma_i = 1 - 2 b_i$. We defer the details to appendix \ref{quboIntoIsing} and simply give here the final expression of the \textit{problem Hamiltonian}:
\begin{align}
    H_P = - \sum_{i} h_i \sigma_i - \sum_{i < j} J_{ij} \sigma_i \sigma_j,
    \label{isingHamilt}
\end{align}
expressed with classical Ising spins $\sigma_i \in \{+1 \equiv \,\uparrow, -1 \equiv\, \downarrow\}$ and the parameters:
\begin{align}
\begin{dcases}
    h_i &= - w_i \left(\bm{c}_i \cdot \tilde{\bm{y}}\right) \\
    J_{ij} &= -\frac{1}{2} (w_i w_j) \, \left(\bm{c}_i \cdot \bm{c}_j\right)
    \end{dcases},
    \label{isingParameters}
\end{align}
where the signal $\tilde{\bm{y}}$ is defined by:
\begin{align}
    \tilde{\bm{y}} = \bm{y} - \frac{1}{2} \sum_{j=1}^N w_j \bm{c}_j.
    \label{yTildeDef}
\end{align}
In terms of Ising variables, an inactive (resp. active) user corresponds to $\sigma_i = \,\uparrow$ (resp. $\sigma_i = \,\downarrow$). The MAP estimator is the ground state of this Hamiltonian. It is the spin configuration $\bm{\sigma}^{(0)} \in \{\uparrow,\downarrow\}^N$ associated to $\bm{b}^{(0)}$ that minimizes the energy of the system. The decodability constraint previously mentioned about the codes $\bm{c}_i$ ensures that the ground state is unique in the idealized situation $\bm{w} = \bm{1}_N$ and $\xi = 0$. 

Let us quickly mention a simple sanity check that confirms that this Hamiltonian is well suited to our problem. We adopt the scenario $\bm{w} = \bm{1}_N$, $\xi = 0$ and relax the NOMA constraint by considering orthogonal identifications sequences, ie. $\bm{c}_i \cdot \bm{c}_j = 0$ for $i \ne j$. Then, the activity pattern is trivially recovered by:
\begin{align}
    b^{(0)}_i = |\bm{y} \cdot \bm{c}_i|.
\end{align}
On the other hand, Eq. \ref{isingParameters} in this ideal situation reduces to:
\begin{align}
\begin{dcases}
    h_i &= ||\bm{c}_i||^2 \sigma_i^{(0)} \\
    J_{ij} &= 0
    \end{dcases},
\end{align}
where $\bm{\sigma}^{(0)}$ is the spin configuration associated to the initial activity pattern $\bm{b}^{(0)}$. The ground state of $H_P$ is trivially $\bm{\sigma}^{(0)}$, as one might reasonably expect. 

The ground state of an Ising Hamiltonian can be found with perfect accuracy through an exhaustive search in the space of configurations $\{\uparrow, \downarrow\}^N$. However, such task would require $\mathcal{O}\left(2^N\right)$ iterations which is why it is known to be NP-hard. In order to perform ground state searching with QA, we promote the Ising spins $\sigma_i$ to quantum spins $\hat{\sigma_i^z}$ \footnote{Recall that $\hat{\sigma_i^{\alpha}}$ denotes the Pauli matrices acting on the $i$-th qubit, where $\alpha = x,y,z$}. The problem Hamiltonian is thus encoded in the operator:
\begin{align}
    \hat{H}_P = - \sum_{i} h_i \hat{\sigma_i^z} - \sum_{i < j} J_{ij} \hat{\sigma_i^z} \hat{\sigma_j^z}.
    \label{quantumIsingH}
\end{align}

\section{State of the art for QA scheduling}
\label{sec:reviewQA}

\begin{figure*}
\centering
    \subfloat[Eigenvalues $\varepsilon_0$ and $\varepsilon_1$ \label{subFig:energyLevelsN8}]{%
    \includegraphics[scale=0.35]{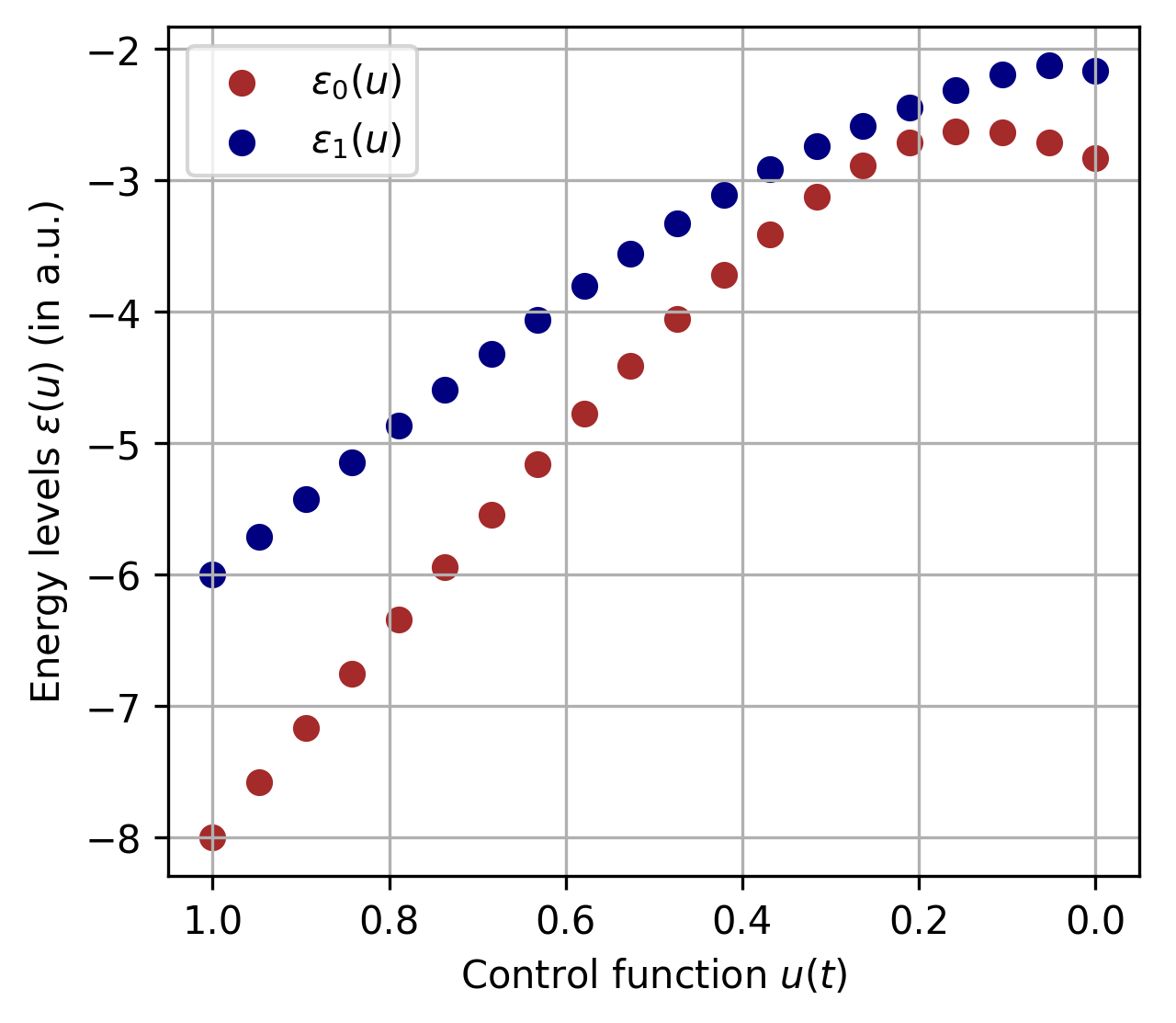}}\hspace{0.5cm}
    \subfloat[Evolution of $\Delta^2$ \label{subFig:gapN8}]{%
    \includegraphics[scale=0.35]{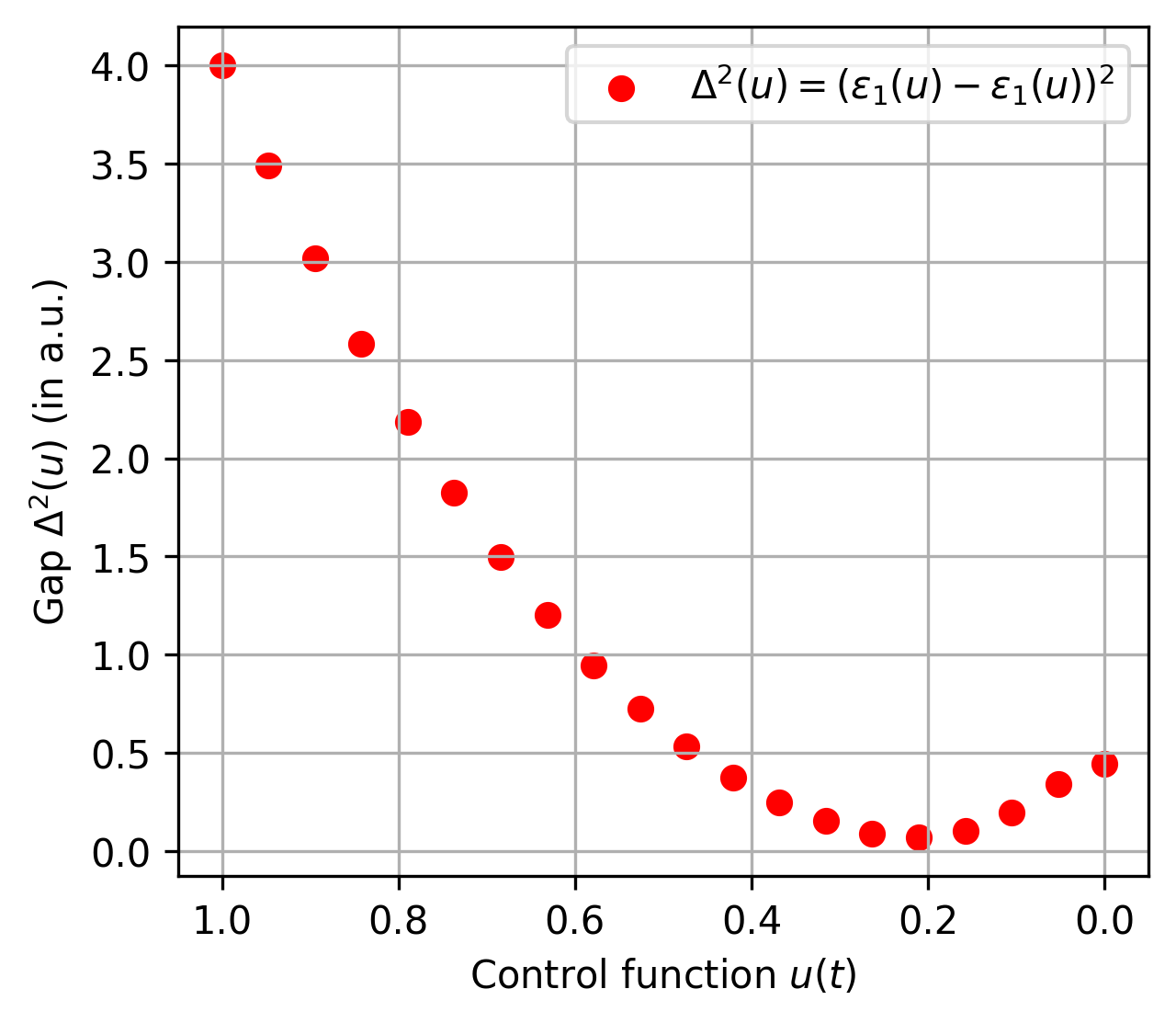}}\hspace{0.5cm}
    \subfloat[$u_{(\bm{y},\bm{w})}(t)$ \label{subFig:exControlFct}]{%
    \includegraphics[scale=0.35]{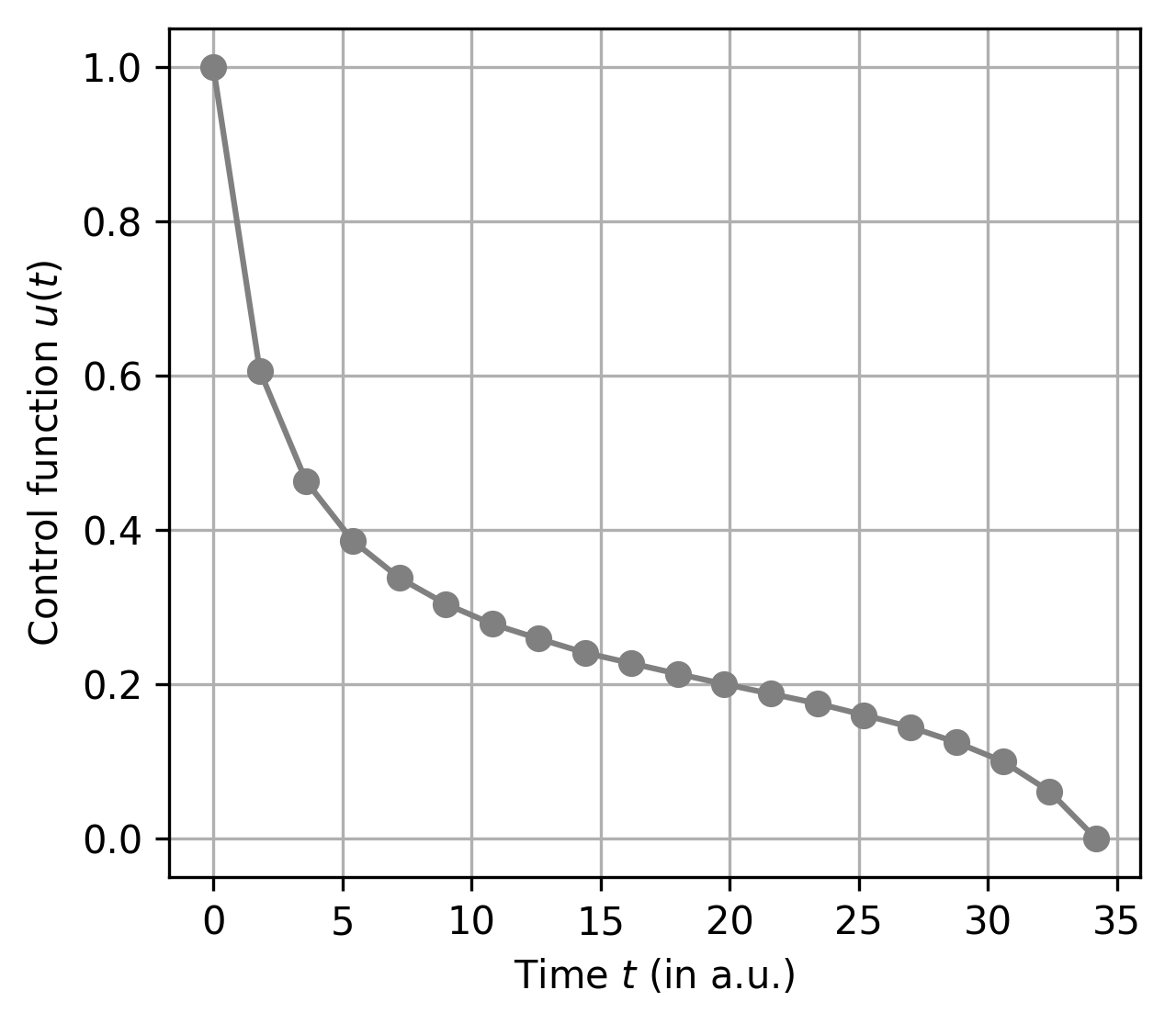}}\hspace{0.5cm}
    \subfloat[$P_{\text{QA}}(t)$ \label{subFig:exPqa}]{%
    \includegraphics[scale=0.35]{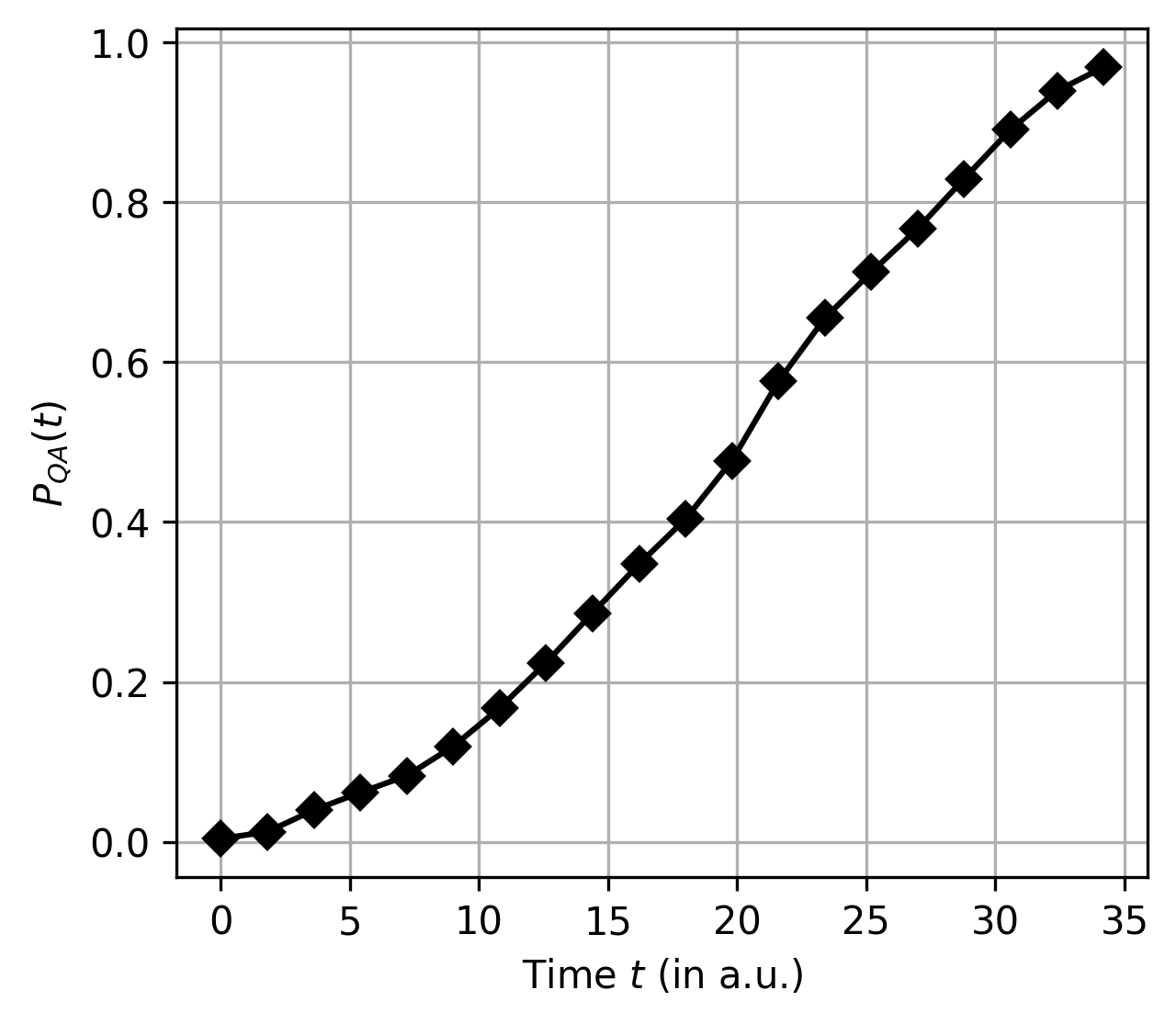}}\hspace{0.5cm}
  \caption{Two first eigenvalues of $\hat{H}(u)$ for some values of the control function $u \in [0,1]$ (a) and corresponding evolution of the spectral gap (b). It allows to solve numerically Eq. \ref{ODEuOptimal} to obtain $u_{(\bm{y},\bm{w})}$ (c), which yields the following evolution for the overlap $P_{\text{QA}}(t)$ (d)} 
  \label{fig:exampleSolveProblem}    
\end{figure*}
\subsection{Basic principle of QA}

Given a quantum Hamiltonian $\hat{H}_P$ whose ground state is hard to find, one uses a control Hamiltonian $\hat{H}_C$ with a simple ground state to initialize the system in this known configuration. In the initial formulation of QA by Kadowaki and Nishimori \cite{kadowaki_quantum_1998}, $\hat{H}_C$ is chosen as the uniform transverse field Hamiltonian:
\begin{align}
    \hat{H}_C = - \sum_{i=1}^N \hat{\sigma_i^x},
    \label{transverseHc}
\end{align}
whose ground state is the uniform superposition $\ket{+}^{\otimes N}$. The QA process slowly evolves the system from an initial dynamic dictated by $\hat{H}_C$ towards a final dynamic encoded by $\hat{H}_P$ through the global Hamiltonian:
\begin{align}
    \hat{H}(t) = (1-u(t)) \hat{H}_P + u(t) \hat{H}_C.
    \label{globalHamilt}
\end{align}

The function $u(t)$ is a control function obeying the initial condition $u(0) = 1$ so that the system starts in the ground state $\ket{+}^{\otimes N}$. If the control function decreases slowly towards 0 during a period $T$, the final state of the system is expected to be the ground state of $\hat{H}_P$ with high probability. This period $T$ is called the annealing time and its duration can be estimated thanks to the adiabatic theorem.

\subsection{Adiabatic approximation}

The so-called adiabatic theorem is a mathematical statement on the time evolution of the global Hamiltonian $\hat{H}(t)$ to ensure that the instantaneous state $\ket{\psi(t)}$ of the system remains close to the instantaneous ground state $\ket{\phi_0(t)}$. Several formulations exist with different precision levels reviewed in \cite{albash_adiabatic_2018, morita_mathematical_2008}. If one denotes $\{\varepsilon_n(t), \ket{\phi_n(t)}\}$ the instantaneous eigensystem of $\hat{H}(t)$ and $\Delta(t) = \varepsilon_1(t) - \varepsilon_0(t)$ is the spectral gap between the ground state and the first excited state, the most commonly used adiabatic approximation reads \cite{hauke_perspectives_2020}:
\begin{align}
    \frac{ |\langle \phi_1(t) | \partial_t \hat{H}(t) | \phi_0(t) \rangle |}{\Delta(t)^2} & \ll 1 \,\,\, \forall t \in [0, T] \Rightarrow P_{\text{QA}}(T) \approx 1,
    \label{AdiaThm}
\end{align}
where we introduced the notation used by \cite{kadowaki_quantum_1998}:
\begin{align}
    P_{\text{QA}}(t) = |\langle \phi_0(T) | \psi(t) \rangle |^2.
\end{align}
The success probability of QA naturally corresponds to $P_{\text{QA}}(T)$. In our case, the time dependence of $\hat{H}(t)$ is fully contained in the control function $u$. Using Eq. \ref{globalHamilt}, the adiabatic approximation reads:
\begin{align}
    \left| \frac{du}{dt} \right| \frac{ |\langle \hat{H}_C - \hat{H}_P \rangle_{1,0}(u) |}{\Delta(u)^2} \ll 1 \,\, \forall t \in [0,T] \,\, \Rightarrow P_{\text{QA}}(T) \approx 1.
    \label{localAdiaThmWithU}
\end{align}
which constraints the control function $u$. It is quite easy to show that the numerator $|\langle \hat{H}_C - \hat{H}_P \rangle_{1,0}(u) |$ scales at most as $\mathcal{O}(\text{Poly}(N))$ (see \cite{morita_mathematical_2008} for instance). Thus, a qualitative discussion of the adiabatic approximation as a function of the system size is often focused on the spectral gap $\Delta^2$. At the end, one usually uses the following constraint on the control function $u$:
\begin{align}
    \left| \frac{du}{dt} \right| \Delta(u)^{-2} \ll 1 \,\, \forall t \in [0,T].
    \label{constraintOnU}
\end{align}

\subsection{Linear scheduling}

Several theoretical analysis of QA use a linear control function \cite{farhi_quantum_2000} to parameterize the annealing process:
\begin{align}
    u_{\text{lin}}(t) = 1 - \frac{t}{T_{\text{lin}}}.
    \label{linearControlFunction}
\end{align}
In this case, $T_{\text{lin}}$ naturally corresponds to the annealing time. However, the first derivative of $u$ is constant which heavily constraints $T_{\text{lin}}$. Indeed, Eq. \ref{constraintOnU} immediately imposes:
\begin{align}
    T_{\text{lin}} \gg \Delta^{-2}_{\text{min}},
    \label{conditionTlin}
\end{align}
where $\Delta^2_{\text{min}} = \underset{0 \leq u \leq 1}{\text{min }}\Delta^2(u)$. However, if the system encounters a first order quantum phase transition during the annealing process, it is known that the minimum of the gap closes exponentially with the system size \cite{hauke_perspectives_2020}. Concretely, it means that $T_{\text{lin}} \sim e^{\alpha N}$ with $\alpha > 1$ so QA would offer no advantage over an exhaustive search to compute $\hat{\bm{b}}^{\text{MAP}}$.

The work \cite{roland_quantum_2002} underlines that in this case one can enhance the annealing time by looking for a control function with a non constant derivative.

\subsection{Optimized control function}

Recall that $u$ is a decreasing function from $u(0) = 1$ to $u(T) = 0$, which means that $du/dt < 0$. As it is done in \cite{roland_quantum_2002}, we can fix an arbitrary small constant $\epsilon > 0$ and impose:
\begin{align}
    \begin{dcases}
        \frac{du}{dt} &= - \epsilon \Delta^2(u(t)) \\
        u(0) &= 1
    \end{dcases},
    \label{ODEu}
\end{align}
which directly comes from Eq. \ref{constraintOnU}. We refer to the solution of ordinary differential equation (ODE) as the optimized control function. Qualitatively, this equation tells that the larger the gap, the faster the control function can decrease. Integrating both sides of this ODE from $t=0$ to $t=T$ yields the same expression than \cite{monroe_programmable_2021} for the annealing time:
\begin{align}
    T = \frac{1}{\epsilon} \int_{0}^1 \frac{du}{\Delta^2(u)}.
    \label{intAnnealingTime}
\end{align}
These two equations \ref{ODEu}, \ref{intAnnealingTime} underline that once the gap is known, one can fully determine the annealing time and the control function associated to a problem encoded in $\hat{H}_P$.

\subsection{An example}

Going back to AUD, one can build the quantum Ising Hamiltonian \ref{quantumIsingH} once the received signal $\bm{y}$ and the channel coefficients $\bm{w}$ are known. Thus, we call a pair $(\bm{y}, \bm{w})$ a \textit{problem instance} that allows to compute the Ising parameters of Eq. \ref{isingParameters}. Since the time dependence of the full Hamiltonian $\hat{H}$ is entirely contained in the control function, one can take $u \in [0,1]$ as an affine parameter to evaluate the spectral gap as a function of $u$ before even knowing the time dependence $u(t)$. Thus, we denote $\Delta^2_{(\bm{y}, \bm{w})}(u)$ the spectral gap associated to the problem instance $(\bm{y}, \bm{w})$. From Eq. \ref{ODEu} and \ref{intAnnealingTime}, one defines the associated control function:
\begin{align}
    \begin{dcases}
        \frac{du_{(\bm{y}, \bm{w})}}{dt} &= - \epsilon \Delta^2_{(\bm{y}, \bm{w})}(u_{(\bm{y}, \bm{w})}(t)) \\
        u_{(\bm{y}, \bm{w})}(0) &= 1
        \label{ODEuOptimal}
    \end{dcases},
\end{align}
which is the optimal one for $(\bm{y}, \bm{w})$. The associated annealing time reads:
\begin{align}
    T_{(\bm{y},\bm{w})} = \frac{1}{\epsilon} \int_{0}^1 \frac{du}{\Delta^2_{(\bm{y}, \bm{w})}(u)}.
\end{align}

As an example, we considered a network of $N=8$ users and generated a problem instance $(\bm{y},\bm{w})$ in the following scenario:
\begin{align*}
    \begin{dcases}
        \text{Activity pattern: }& \bm{b}^{(0)} = (1,0,0,0,0,0,0,0) \\
        \text{Perfect channels: }& \bm{w} \sim \text{Dirac}(\bm{1}_N) \\
        \text{No additive noise: }& \xi = 0
    \end{dcases}.
\end{align*}
We show on Fig. \ref{subFig:energyLevelsN8} the two first energy levels of the global Hamiltonian $\hat{H}(u)$. Fig. \ref{subFig:gapN8} gives the corresponding spectral gap $\Delta^2_{(\bm{y},\bm{w})}$. These results have been obtained by exact diagonalization of the Hamiltonian using the QuTiP\footnote{\url{https://qutip.org}} library. Then we numerically evaluate the annealing time $T_{(\bm{y},\bm{w})}$ and solve Eq. \ref{ODEuOptimal} on the interval $[0,T_{(\bm{y},\bm{w})}]$ to obtain the optimized control function (Fig. \ref{subFig:exControlFct}). To do so, we fixed our precision parameter to $\epsilon = 0.1$. The evolution of the overlap during the annealing process scheduled with this control function is finally shown on Fig. \ref{subFig:exPqa}. One can see that the success probability $P_{\text{QA}}\left(T_{(\bm{y},\bm{w})}\right)$ reached at the end of the evolution is very close to 1, as expected.
However, it would be very long and inconvenient to require to compute $T_{(\bm{y},\bm{w})}$ and $u_{(\bm{y},\bm{w})}$ for each instance $(\bm{y},\bm{w})$ of our problem. The goal of the following sections is to emphasize the possibility to build a generic approach to parameterize a QA process for AUD.

\section{Proposition: generic control function with perfect channels}
\label{sec:controlFctNoChannel}
We would like to check if, despite the randomness of the parameters $(\bm{y}, \bm{w})$, we can estimate a generic control function for a given size of the network $N$. We first consider a simple scenario with perfect channels.

\subsection{Mean gap in the absence of additive noise}

\begin{figure}
    \centering
    \includegraphics[scale=0.4]{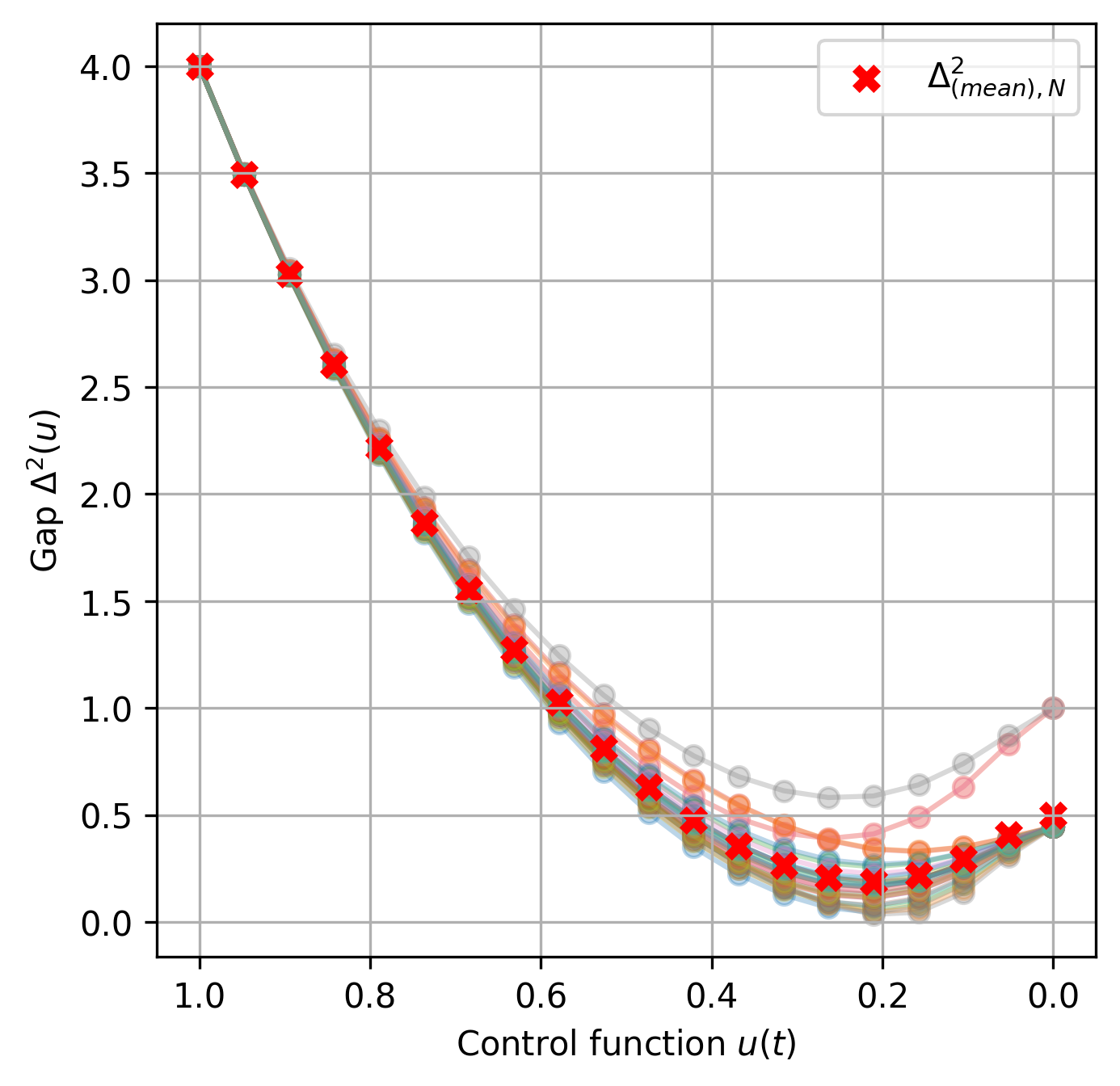}
    \caption{Several gaps $\Delta^2_{(\bm{y},\bm{w})}$ associated to random samples of $(\bm{y},\bm{w} = \bm{1}_N)$ and their mean value (in red) for $N=8$.}
    \label{fig:gapMultiTestN8}
\end{figure}

Let us begin with a scenario where the size of the network $N$ is fixed and where the additive Gaussian noise is neglected, which amounts to fix $\xi = 0$. We randomly generated 30 samples of $(\bm{y}, \bm{w})$ to check how the associated gaps $\Delta^2_{(\bm{y},\bm{w})}$ behave. The signals $\bm{y}$ are generated from uniform samples of activity patterns $\bm{b}^{(0)}$ and the channel coefficients are given by $\bm{w} = \bm{1}_N$. Interestingly, Fig. \ref{fig:gapMultiTestN8} shows that a pattern emerges for the shape of the gap against $u$ despite the dependence of the local magnetic fields $h_i$ in $\bm{y}$. 

These results motivate the definition of the \textit{mean gap}:
\begin{align}
    \Delta^2_{(\text{mean}),N}(u) = \mathbb{E}_{(\bm{y},\bm{w})}\left(\Delta^2\right).
\end{align}
In our specific case, this quantity can be exactly computed. Indeed, the channel coefficients are fixed to  $\bm{w} = \bm{1}_N$ and the received signal $\bm{y}$ is fully determined from the initial activity pattern $\bm{b}^{(0)}$. Using that $\bm{b}^{(0)}$ is uniformly distributed, one has:
\begin{align}
    \Delta^2_{(\text{mean}),N} = \frac{1}{2^N} \sum_{\bm{b}^{(0)} \in \{0,1\}^N} \Delta^2_{(\bm{y}, \bm{w}=\bm{1}_N)}(u).
    \label{meanGapNoChannel}
\end{align}
Then, we can use this mean gap to compute the associated annealing time $T_{(\text{mean}),N}$ with Eq. \ref{intAnnealingTime}:
\begin{align}
    T_{(\text{mean}),N} = \frac{1}{\epsilon} \int_{0}^1 \frac{du}{\Delta^2_{(\text{mean}),N}(u)},
    \label{intMeanAnnealingT}
\end{align}
and the resulting control function $u_{(\text{mean}),N}(t)$ from Eq. \ref{ODEu}:
\begin{align}
    \begin{dcases}
        \frac{du_{(\text{mean}),N}}{dt} &= - \epsilon \Delta^2_{(\text{mean}),N}\left(u_{(\text{mean}),N}(t)\right) \\
        u_{(\text{mean}),N}(0) &= 1
    \end{dcases}.
    \label{uMeanODE}
\end{align}

\begin{figure}
    \centering
    \includegraphics[scale=0.4]{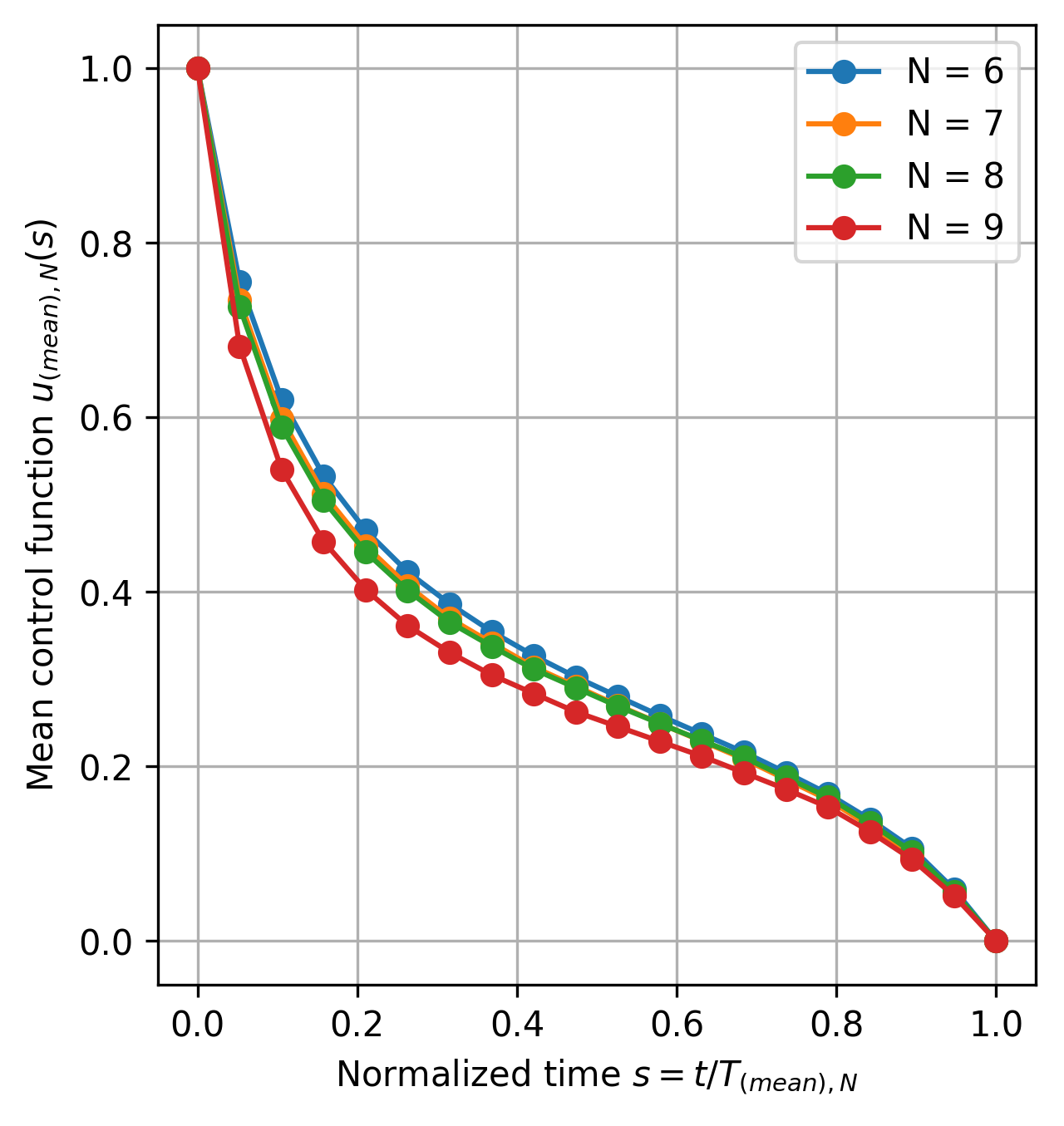}
    \caption{Shape of the mean control functions obtained for $N=6,\dots,9$ with respect to $s = t/T_{(\text{mean}),N}$}
    \label{fig:meanControlFctMutliN}
\end{figure}

Once for all, we adopt the precision level $\epsilon = 0.1$. We reported the shape obtained for $u_{(\text{mean}),N}(s)$ for $N = 6,\dots,9$ on Fig. \ref{fig:meanControlFctMutliN} with respect to the normalized time $s = t/T_{(\text{mean}),N}$. Then, we must test for each value of $N$ whether these functions are good candidates to schedule a QA process on the period $t \in [0, T_{(\text{mean}),N}]$ to estimate $\hat{\bm{b}}^{\text{MAP}}$.

\begin{figure*}
\centering
    \subfloat[$N=6$ \label{subFig:histoMeanGapOptiN6_snr20}]{%
    \includegraphics[scale=0.3]{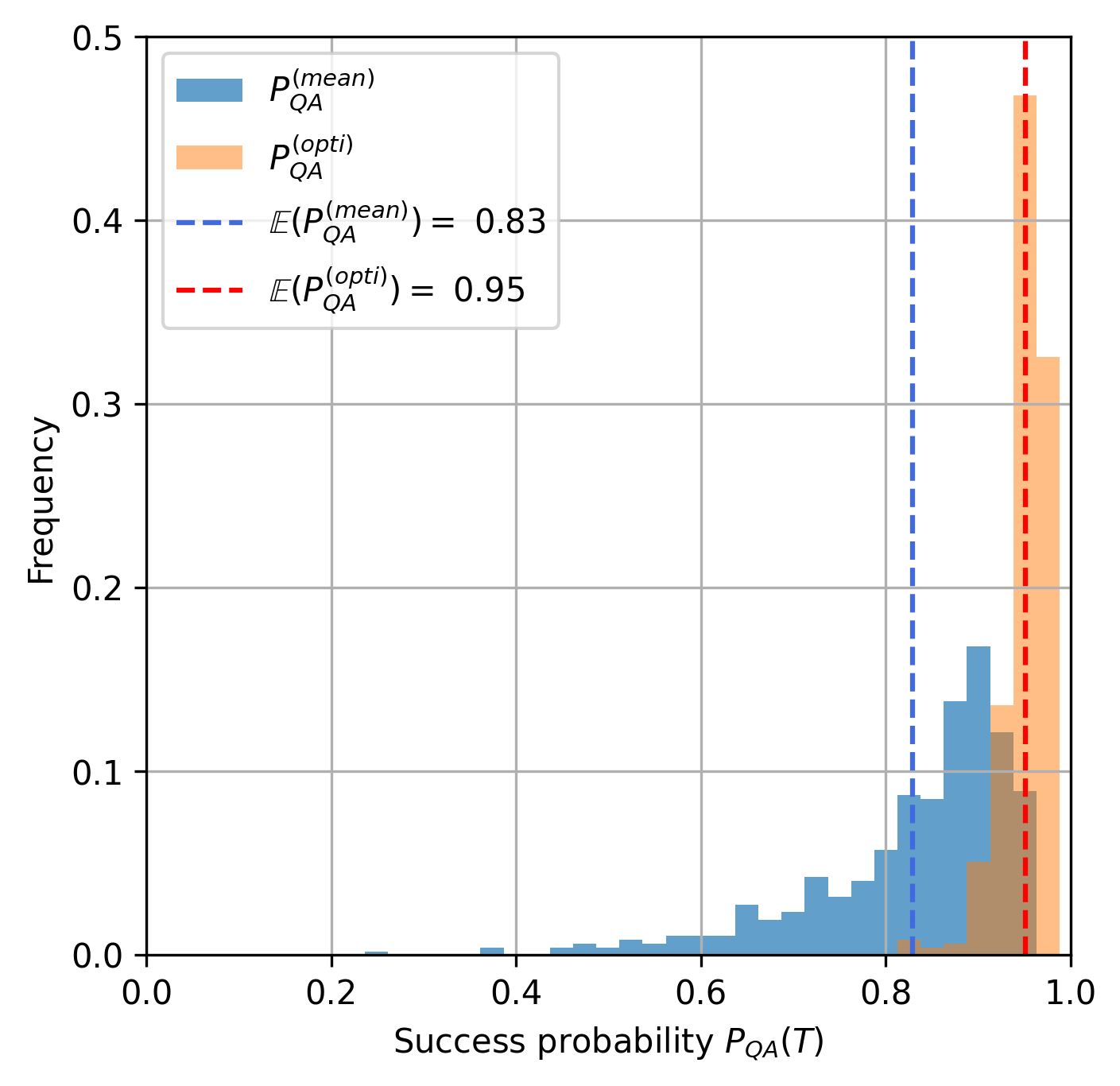}}\hspace{0.5cm}
    \subfloat[$N=7$ \label{subFig:histoMeanGapOptiN7_snr20}]{%
    \includegraphics[scale=0.3]{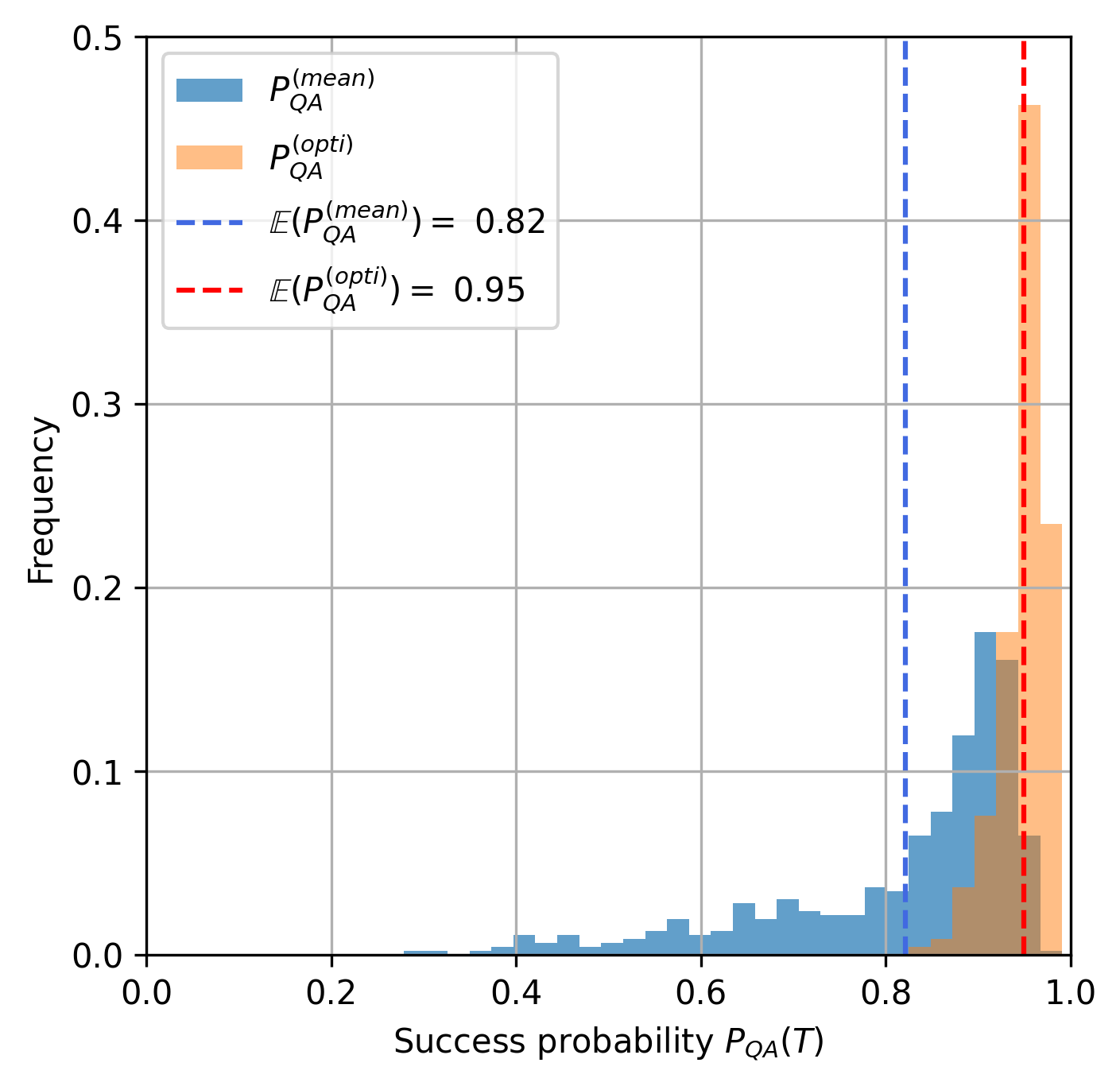}}\hspace{0.5cm}
    \subfloat[$N=8$ \label{subFig:histoMeanGapOptiN8_snr20}]{%
    \includegraphics[scale=0.3]{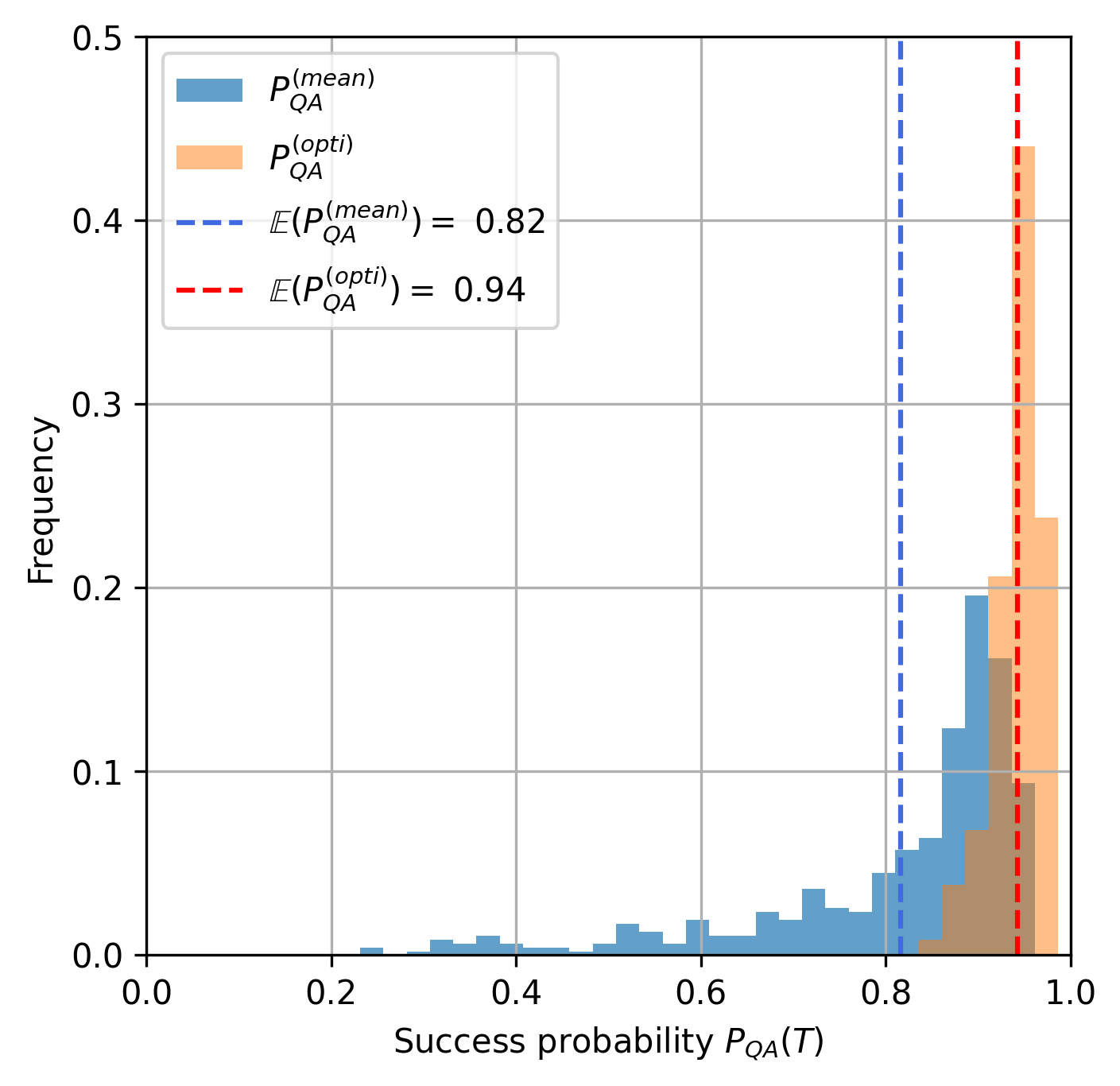}}\hspace{0.5cm}
    \subfloat[$N=9$ \label{subFig:histoMeanGapOptiN9_snr20}]{%
    \includegraphics[scale=0.3]{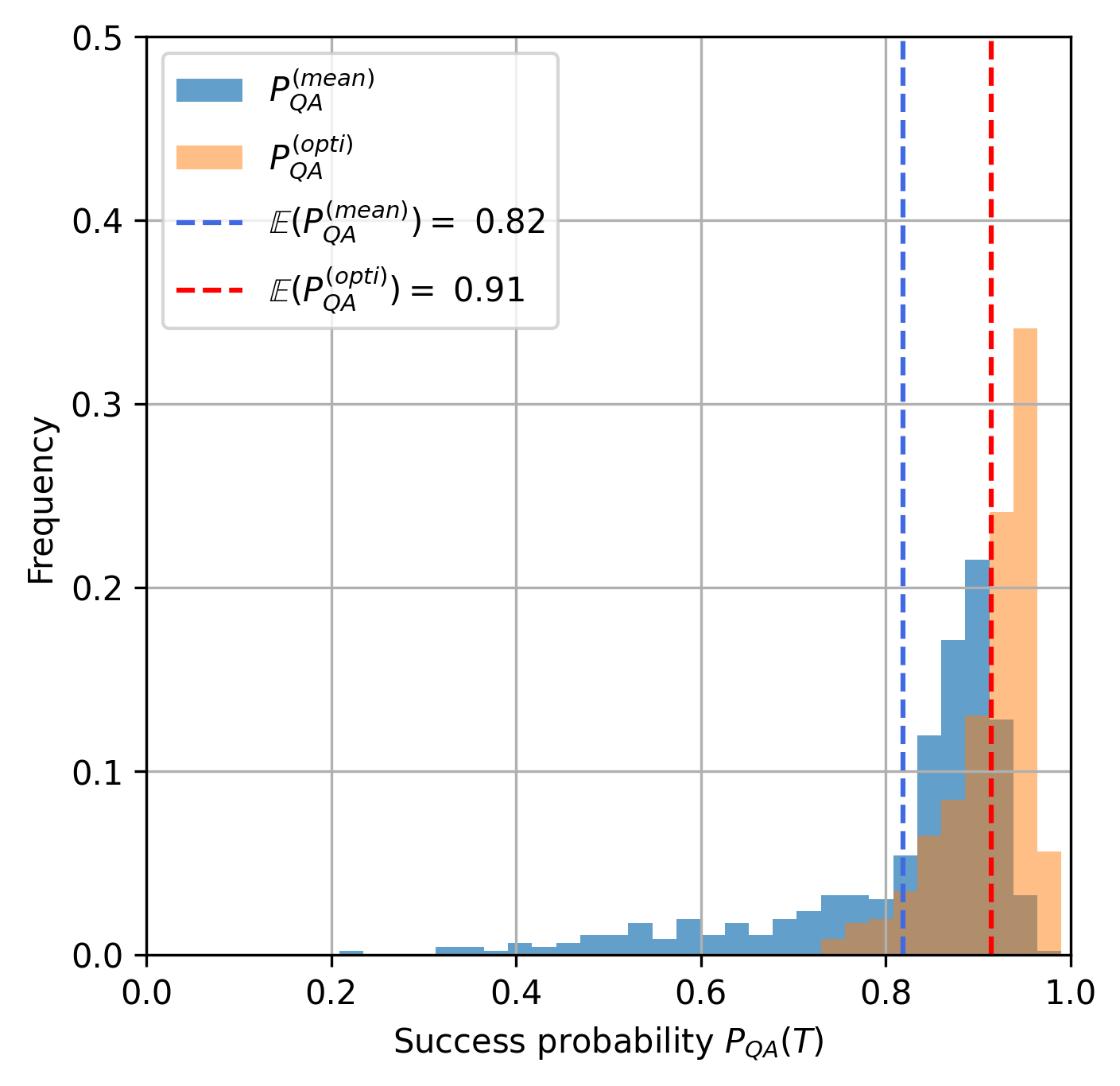}}
  \caption{Distributions of $P_{\text{QA}}^{(\text{mean})}$ (in blue) and $P_{\text{QA}}^{(\text{opti})}$ (in red) for several network sizes $N$. The noise level corresponds to $\text{SNR} = 20 \text{dB}$ We performed $N_{\text{samples}} = 500$ samples for each histogram.} 
  \label{fig:OptiVsMeanNoChannel20db}    
\end{figure*}

\begin{figure*}
\centering
    \subfloat[$N=6$ \label{subFig:histoMeanGapOptiN6_snr15}]{%
    \includegraphics[scale=0.3]{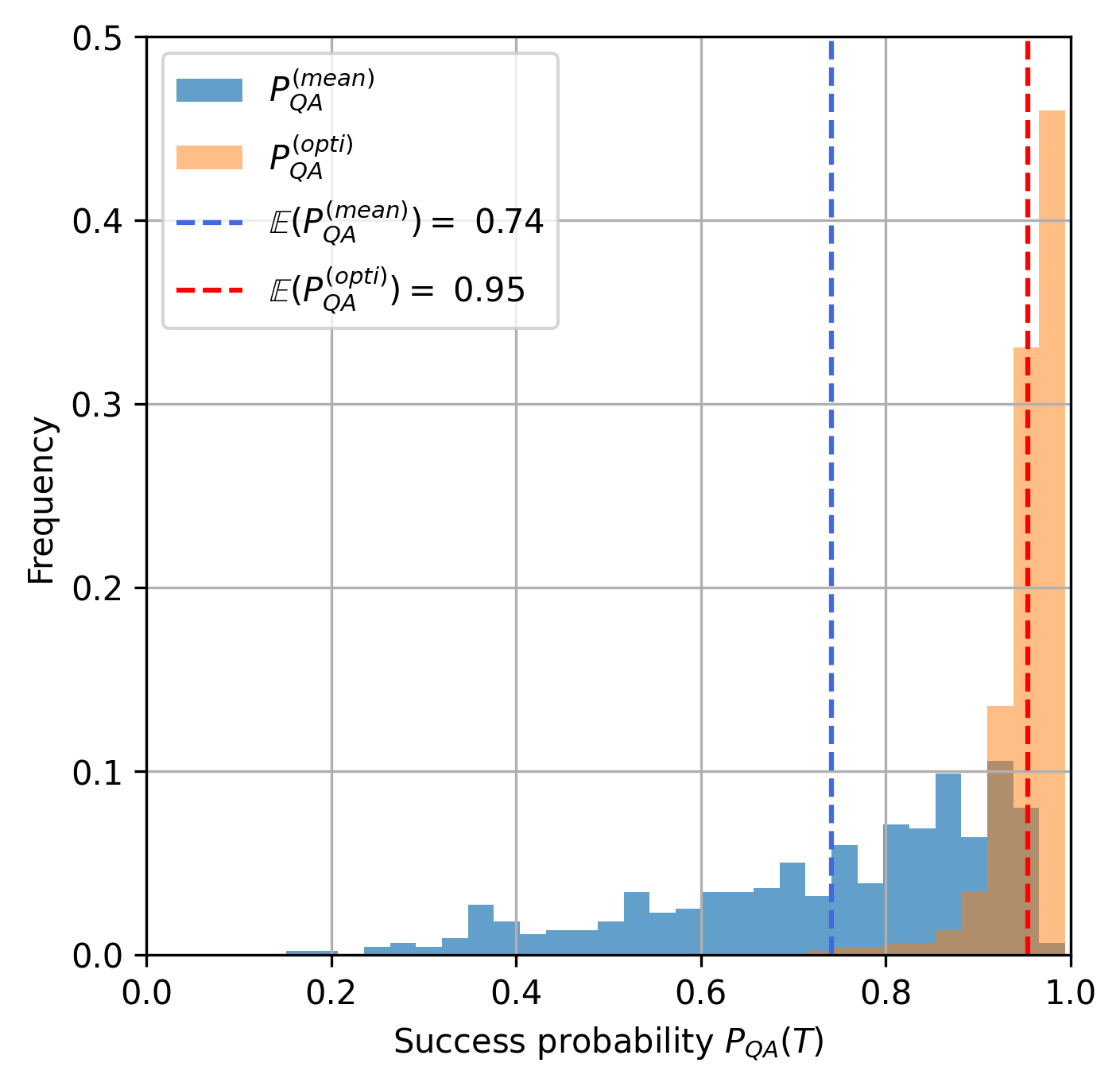}}\hspace{0.5cm}
    \subfloat[$N=7$ \label{subFig:histoMeanGapOptiN7_snr15}]{%
    \includegraphics[scale=0.3]{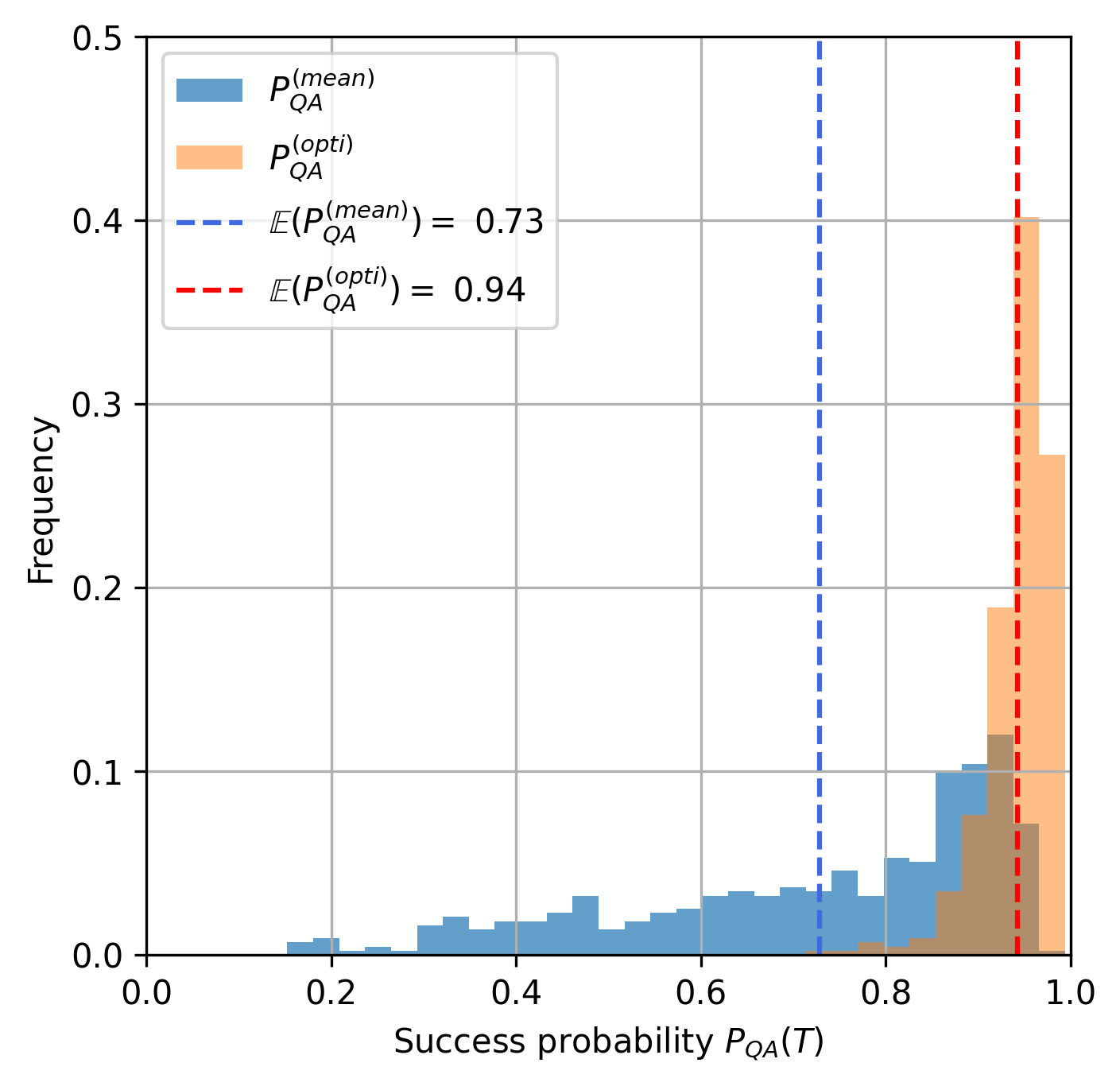}}\hspace{0.5cm}
    \subfloat[$N=8$ \label{subFig:histoMeanGapOptiN8_snr15}]{%
    \includegraphics[scale=0.3]{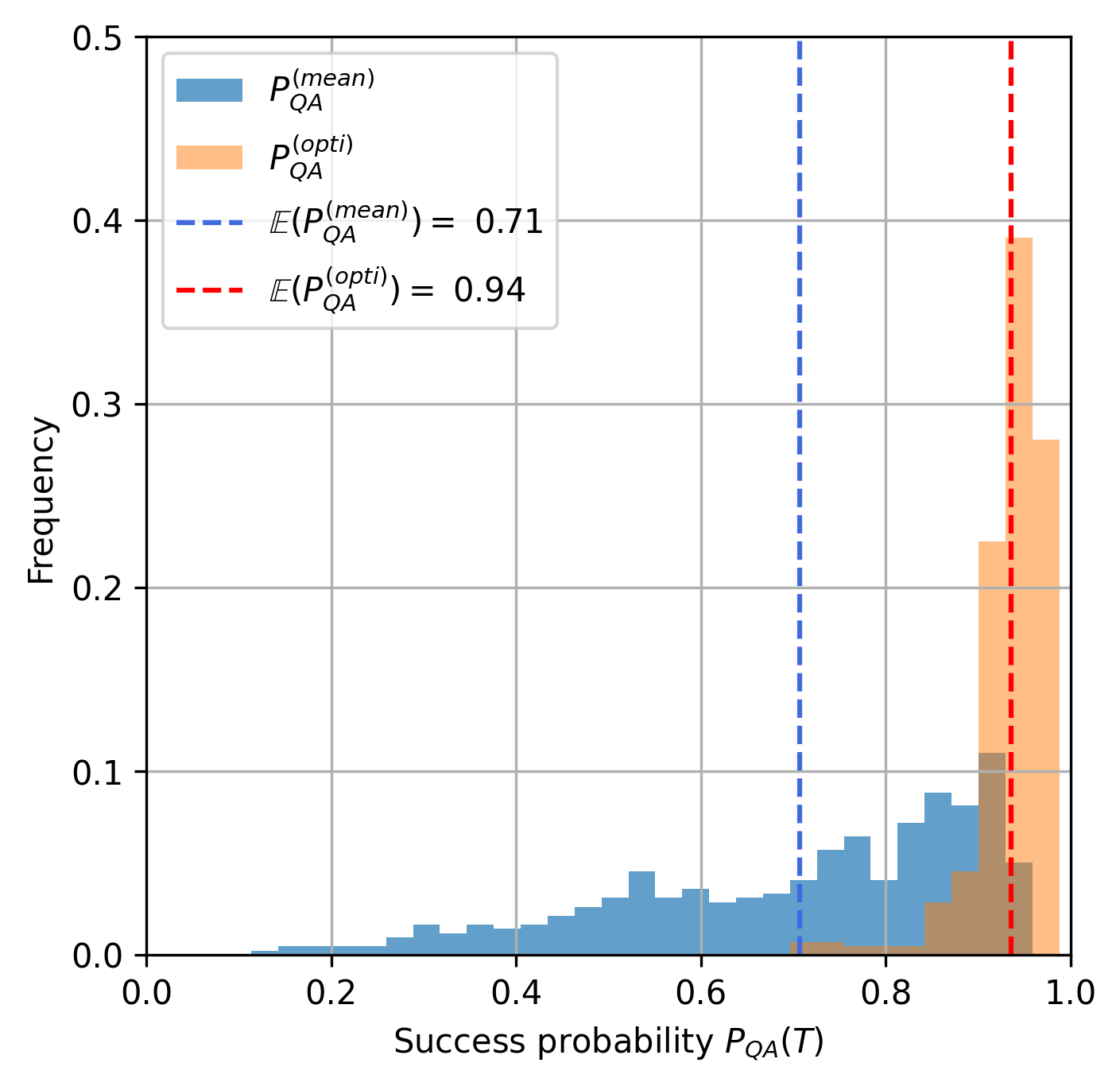}}\hspace{0.5cm}
    \subfloat[$N=9$ \label{subFig:histoMeanGapOptiN9_snr15}]{%
    \includegraphics[scale=0.3]{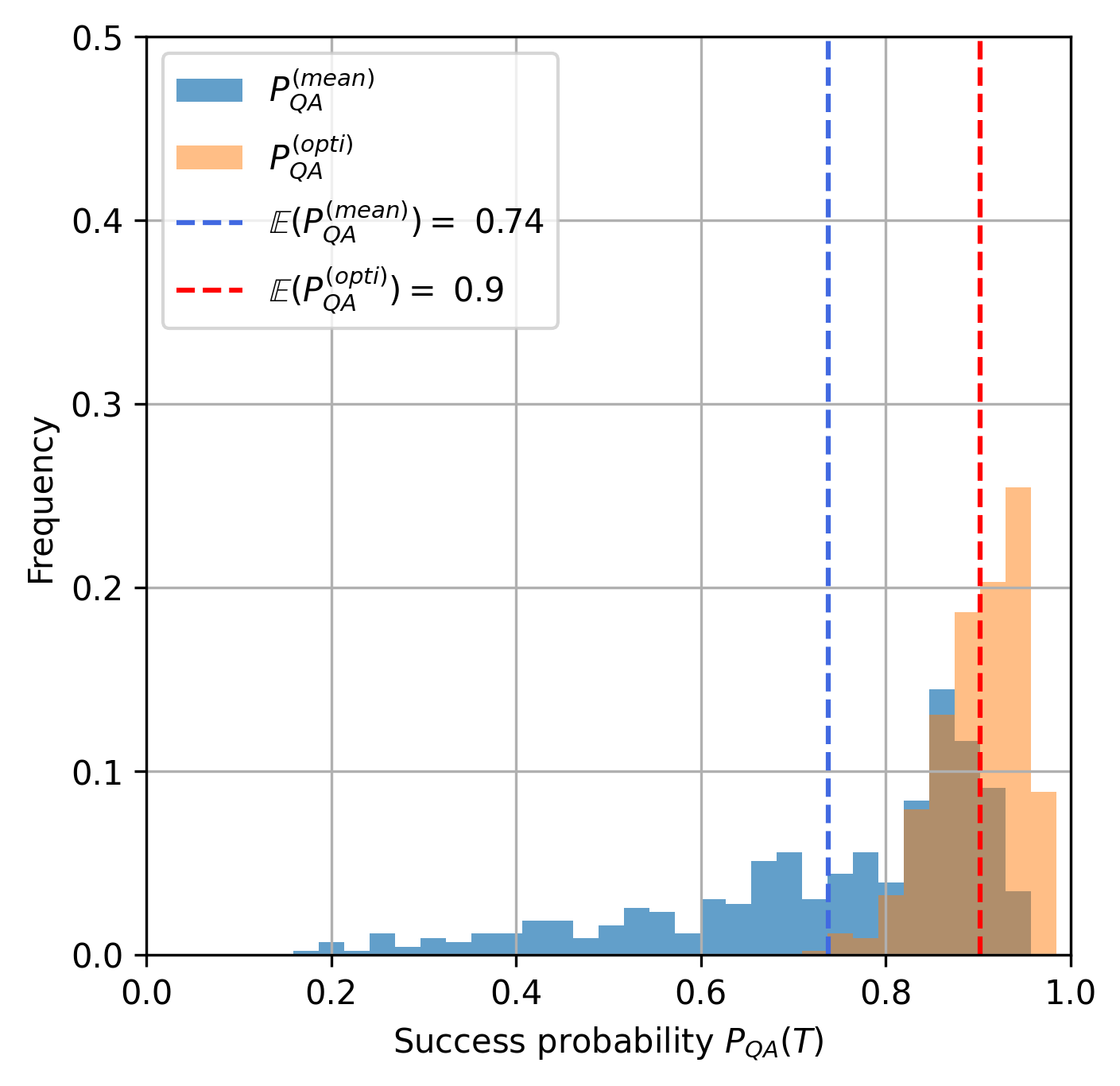}}
  \caption{Same distributions than Fig. \ref{fig:OptiVsMeanNoChannel20db} but with $\text{SNR} = 15\text{dB}$. \hspace{30cm} $ \,\,\,\,\,\,\,\,\,\,\,\,\,\,\,\,\,\,\,\,\,\,\,\,\,\,\,\,\,\,\,\,\,\,\,\,\,\,\,\,\,\,\,\,\,\,\,\,\,\,\,\,\,\,\,\,\,\,$} 
  \label{fig:OptiVsMeanNoChannel15db}    
\end{figure*}

\subsection{Definition of the metrics}

We will compare for several $(\bm{y},\bm{w})$ the success probability of a QA process scheduled with our approach with the one that would been obtained with an optimal scheduling for each problem instance. To do so, one need to define appropriate metrics. We first generate a pair $(\bm{y}, \bm{w})$ and simulate a QA process parameterized with $u_{(\text{mean}),N}$ by numerically solving the Schrödinger's equation:
\begin{align}
    i \partial_t \ket{\psi(t)} = \hat{H}(u_{(\text{mean}),N}(t)) \ket{\psi(t)} \,\, , \,\, t \in [0,T_{(\text{mean}),N}],
    \label{schroEq}
\end{align}
with the initial condition $\ket{\psi(0)} = \ket{+}^{\otimes N}$ (we work with natural units where $\hbar =1)$ using again the QuTiP library. The obtained success probability is denoted with the short-hand notation:
\begin{align}
    P_{\text{QA}}\left(T_{(\text{mean}),N} ; (\bm{y}, \bm{w})\right) \equiv P_{\text{QA}}^{(\text{mean})}(\bm{y}, \bm{w}). 
\end{align}
On the other hand, the reference success probability is the one corresponding to a QA process scheduled with the well-suited optimized control function $u_{(\bm{y},\bm{w})}$ (Eq. \ref{ODEu}) and the corresponding annealing time $T_{(\bm{y},\bm{w})}$ (Eq. \ref{intAnnealingTime}). We adopt also a short hand notation:
\begin{align}
    P_{\text{QA}}\left(T_{(\bm{y},\bm{w})} ; (\bm{y}, \bm{w})\right) \equiv P_{\text{QA}}^{(\text{opti})}(\bm{y}, \bm{w}).
\end{align}

Since the instances $(\bm{y},\bm{w})$ are randomly distributed, $P_{\text{QA}}^{(\text{mean})}$ $P_{\text{QA}}^{(\text{opti})}$ are random variables. Thus, we propose to use their expectation values to quantify the reliability of each scheduling strategy. We turn on the additive Gaussian noise to realize the performance analysis, hence the expectation values are given by:
\begin{align}
\mathbb{E}_{(\bm{y},\bm{w})}\left(P_{\text{QA}}^{(\alpha)}\right) = \sum_{\bm{b}^{(0)} \in \{0,1\}^N} \int d\bm{n} f(\bm{b}^{(0)}, \bm{n}) P_{\text{QA}}^{(\alpha)}(\bm{y},\bm{w}),
\label{defExpValPqa}
\end{align}
for $\alpha \in \{\text{mean}, \text{opti}\}$. The function $f(\bm{b}^{(0)}, \bm{n})$ is the probability density function associated to the instances $(\bm{y}, \bm{w})$. The channel coefficients are still fixed to $\bm{w} = \bm{1}_N$ but the signal $\bm{y}$ is now generated from the sample of an activity pattern $\bm{b}^{(0)}$ and a Gaussian vector $\bm{n}$. Those two parameters are independently distributed, respectively following the laws $\mathcal{U}\left(\{0,1\}^N\right)$ and $\mathcal{N}(0,\xi^2)$. Thus, the weights of Eq. \ref{defExpValPqa} are given by:
\begin{align}
    f(\bm{b}^{(0)}, \bm{n}) = \frac{1}{2^N} \times \frac{1}{(2\pi\xi)^{M/2}} \exp\left(-\frac{\lVert \bm{n} \rVert^2}{2 \xi^2}\right).
    \label{weightsInstanceNoChannel}
\end{align}

We also need a metric to quantify the intensity of the Gaussian noise. A well adapted one is the signal-to-noise ratio (SNR) expressed in dB. Since the power of the signal sent individually by each user is normalized to $P_{\text{signal},i} = 1$, the SNR roughly reads:
\begin{align}
    \text{SNR}  = -20 \log_{10}\left(\xi\right).
\end{align}

\subsection{Reliability of our strategy}

The instances $(\bm{y}, \bm{w})$ have no longer a finite support. Thus, we generated $N_{\text{samples}} = 500$ pairs $(\bm{y}, \bm{w} = \bm{1}_N)$ according to the weights of Eq.\ref{weightsInstanceNoChannel} for $N = 6, \dots,9$ to estimate the distributions of $P_{\text{QA}}^{(\text{mean})}$ and $P_{\text{QA}}^{(\text{opti})}$. The expectation values for $\alpha \in \{\text{mean}, \text{opti}\}$ are estimated by:
\begin{align}
\mathbb{E}_{(\bm{y},\bm{w})}\left(P_{\text{QA}}^{(\alpha)}\right) = \frac{1}{N_{\text{samples}}} \sum_{(\bm{y},\bm{w}) \in \text{samples}} P_{\text{QA}}^{(\alpha)}(\bm{y}, \bm{w}).
\end{align}

We reported on Fig. \ref{fig:OptiVsMeanNoChannel20db} the obtained histograms for a noise level corresponding to $\text{SNR} = 20\text{dB}$. We can observe that the values of $P_{\text{QA}}^{(\text{mean})}$ are more spread out along the axis than the values of $P_{\text{QA}}^{(\text{opti})}$ which is not surprising. Indeed, the control function $u_{\text{mean},N}$ cannot be well suited for all possible outcomes $(\bm{y},\bm{w})$. Nevertheless, the expectation value $\mathbb{E}_{(\bm{y},\bm{w})}\left(P_{\text{QA}}^{(\text{mean})}\right)$ is always above 0.8 which is reasonably below $\mathbb{E}_{(\bm{y},\bm{w})}\left(P_{\text{QA}}^{(\text{opti})}\right)$.

We also reported the obtained distributions for a level of noise corresponding to $\text{SNR} = 15\text{dB}$. The gap between $\mathbb{E}_{(\bm{y},\bm{w})}\left(P_{\text{QA}}^{(\text{mean})}\right)$ and $\mathbb{E}_{(\bm{y},\bm{w})}\left(P_{\text{QA}}^{(\text{opti})}\right)$ is higher but our approach can still offer a mean success probability above 0.7.

Thus, our generic control function appears as a good compromise to schedule a QA approach for any set of parameters $(\bm{y}, \bm{w})$ given a size of the network $N$

\begin{figure*}
\centering
    \subfloat[$N=6$ \label{subFig:histoMeanGapLinN6_snr20}]{%
    \includegraphics[scale=0.3]{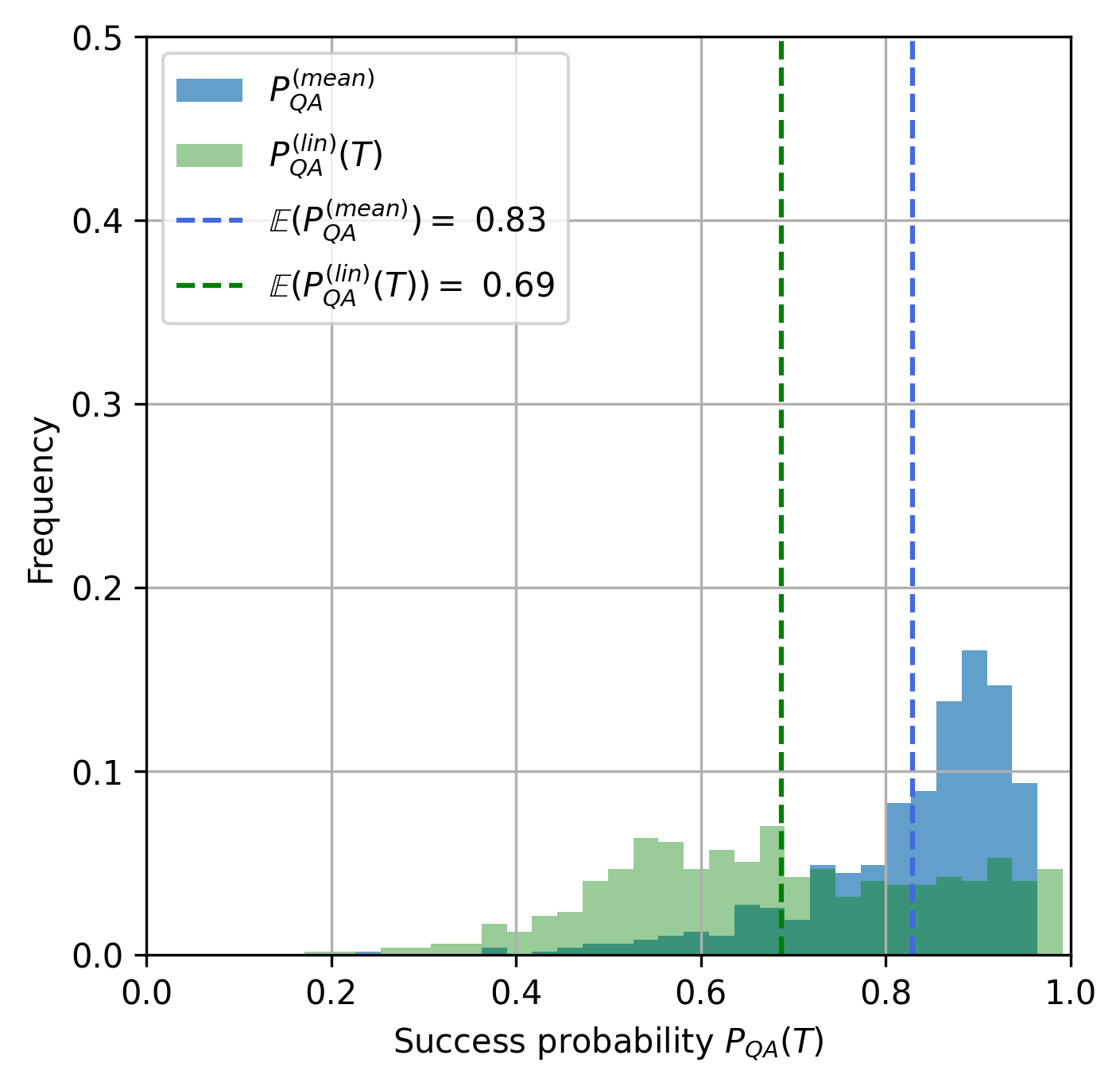}}\hspace{0.5cm}
    \subfloat[$N=7$ \label{subFig:histoMeanGapLinN7_snr20}]{%
    \includegraphics[scale=0.3]{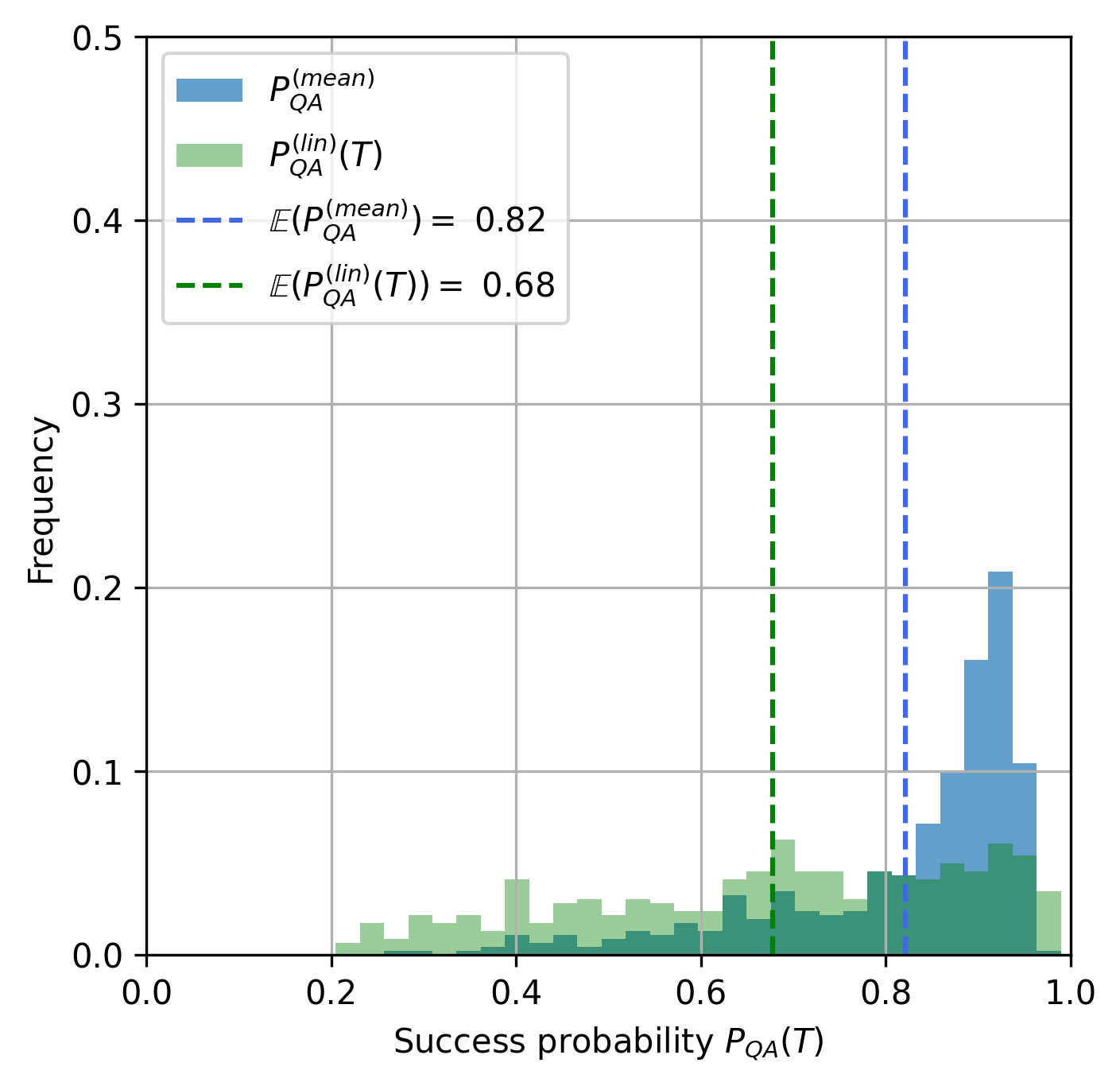}}\hspace{0.5cm}
    \subfloat[$N=8$ \label{subFig:histoMeanGapLinN8_snr20}]{%
    \includegraphics[scale=0.3]{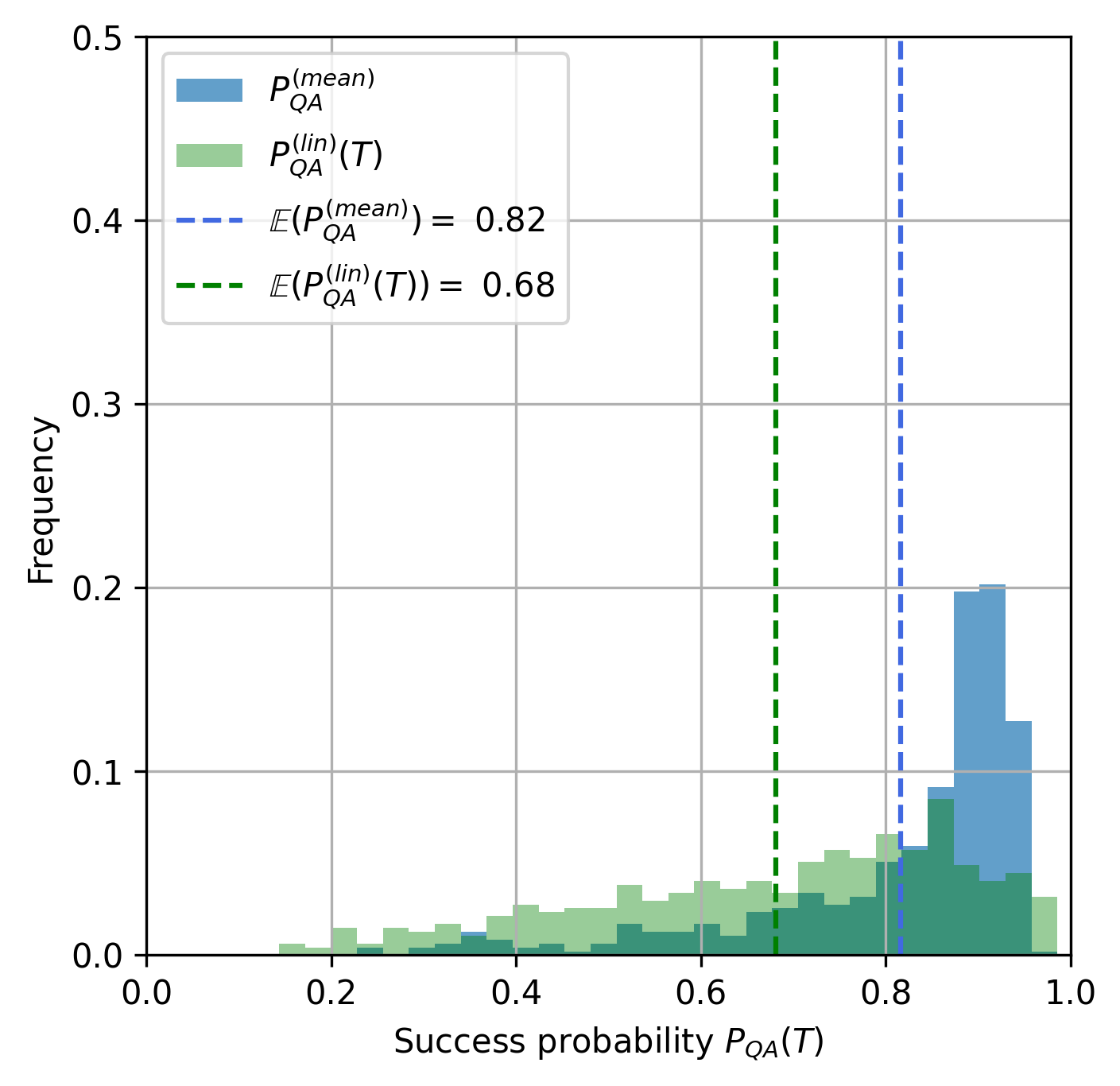}}\hspace{0.5cm}
    \subfloat[$N=9$ \label{subFig:histoMeanGapLinN9_snr20}]{%
    \includegraphics[scale=0.3]{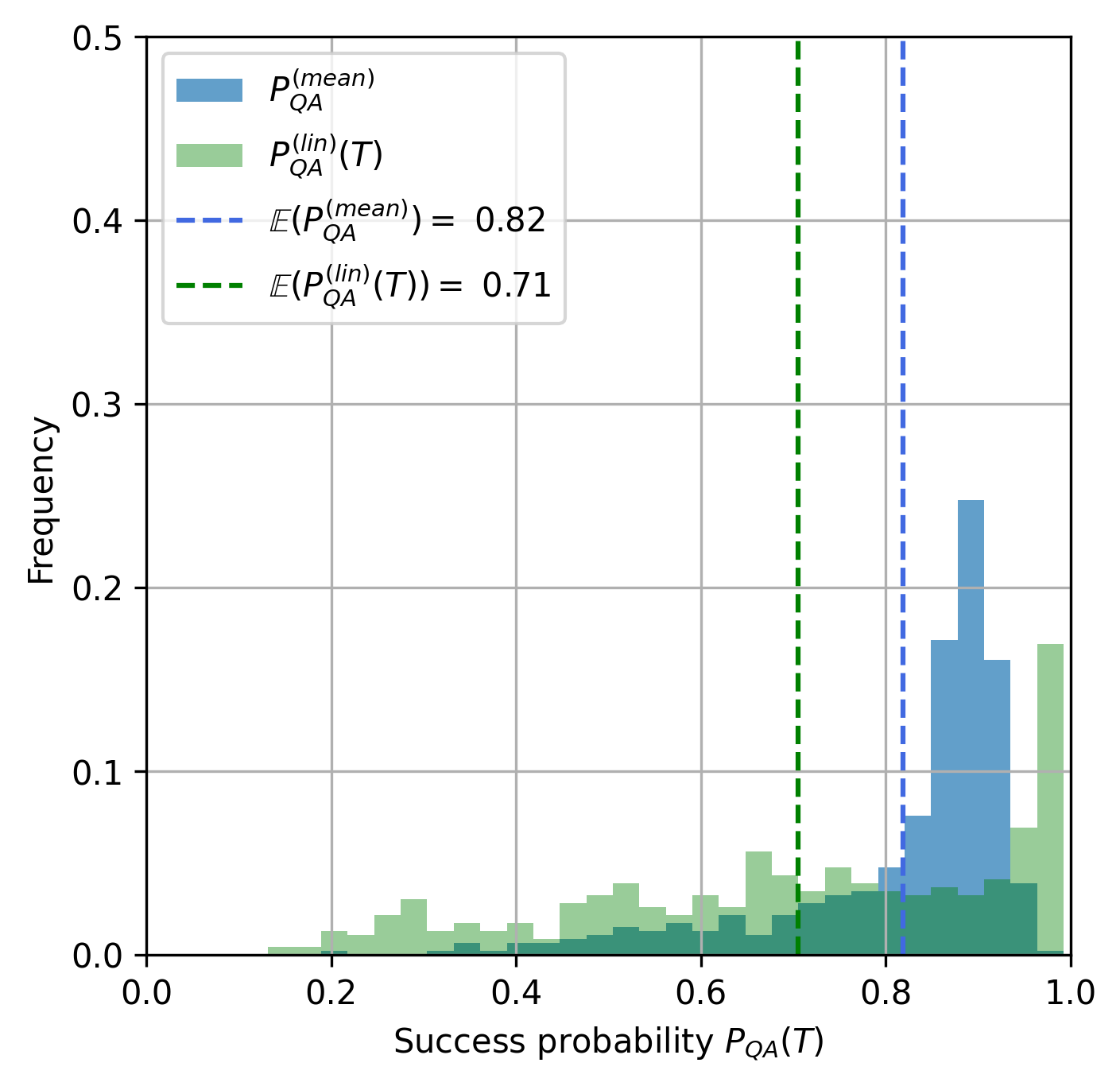}}
  \caption{Distributions of $P_{\text{QA}}^{(\text{mean})}$ (in blue) and $P_{\text{QA}}^{(\text{lin})}$ (in green) for several network sizes $N$. The noise level corresponds to $\text{SNR} = 20 \text{dB}$ We performed $N_{\text{samples}} = 500$ samples for each histogram.} 
  \label{fig:LinVsMeanNoChannel20db}    
\end{figure*}

\begin{figure*}
\centering
    \subfloat[$N=6$ \label{subFig:histoMeanGapLinN6_snr15}]{%
    \includegraphics[scale=0.3]{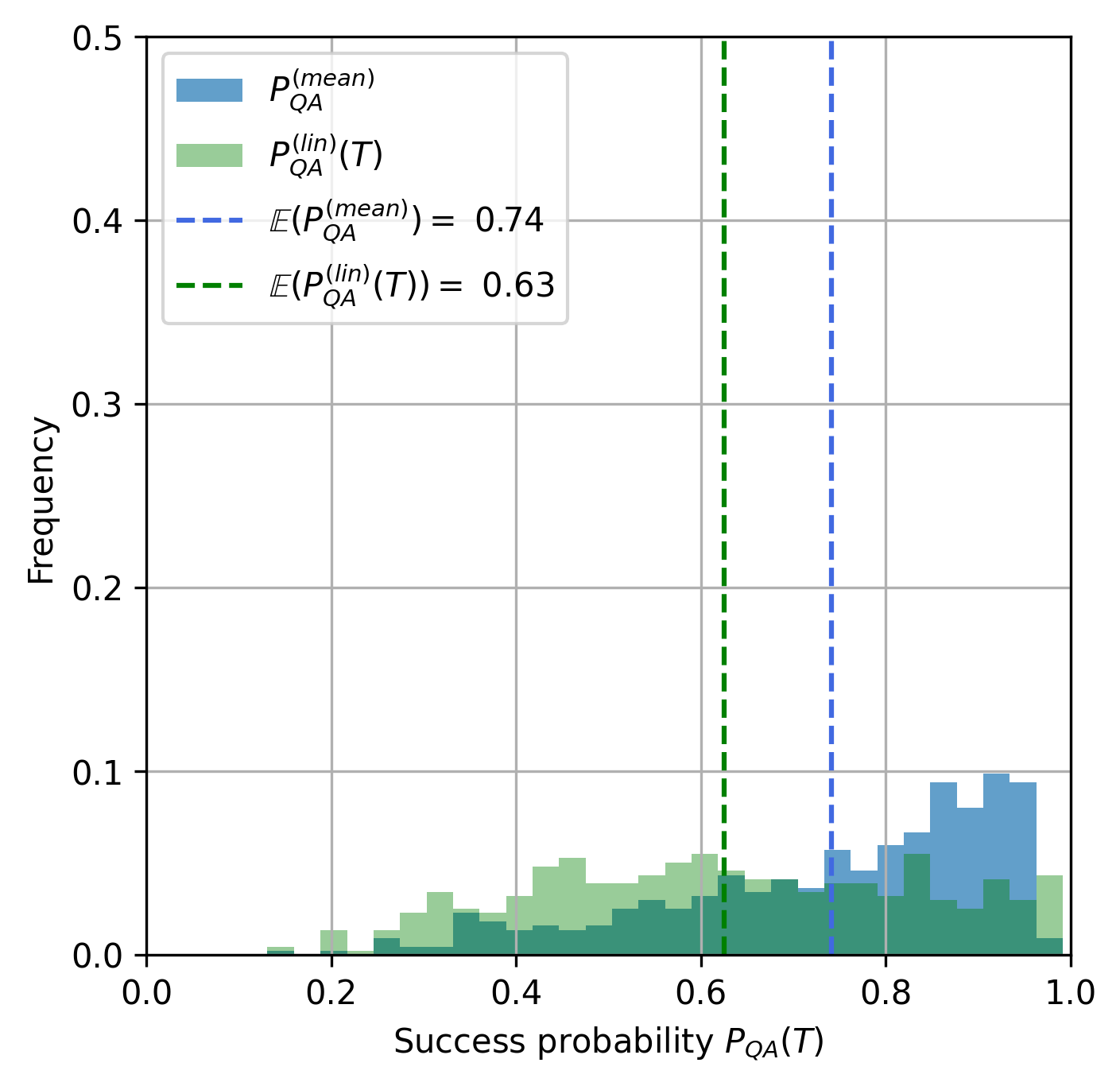}}\hspace{0.5cm}
    \subfloat[$N=7$ \label{subFig:histoMeanGapLinN7_snr15}]{%
    \includegraphics[scale=0.3]{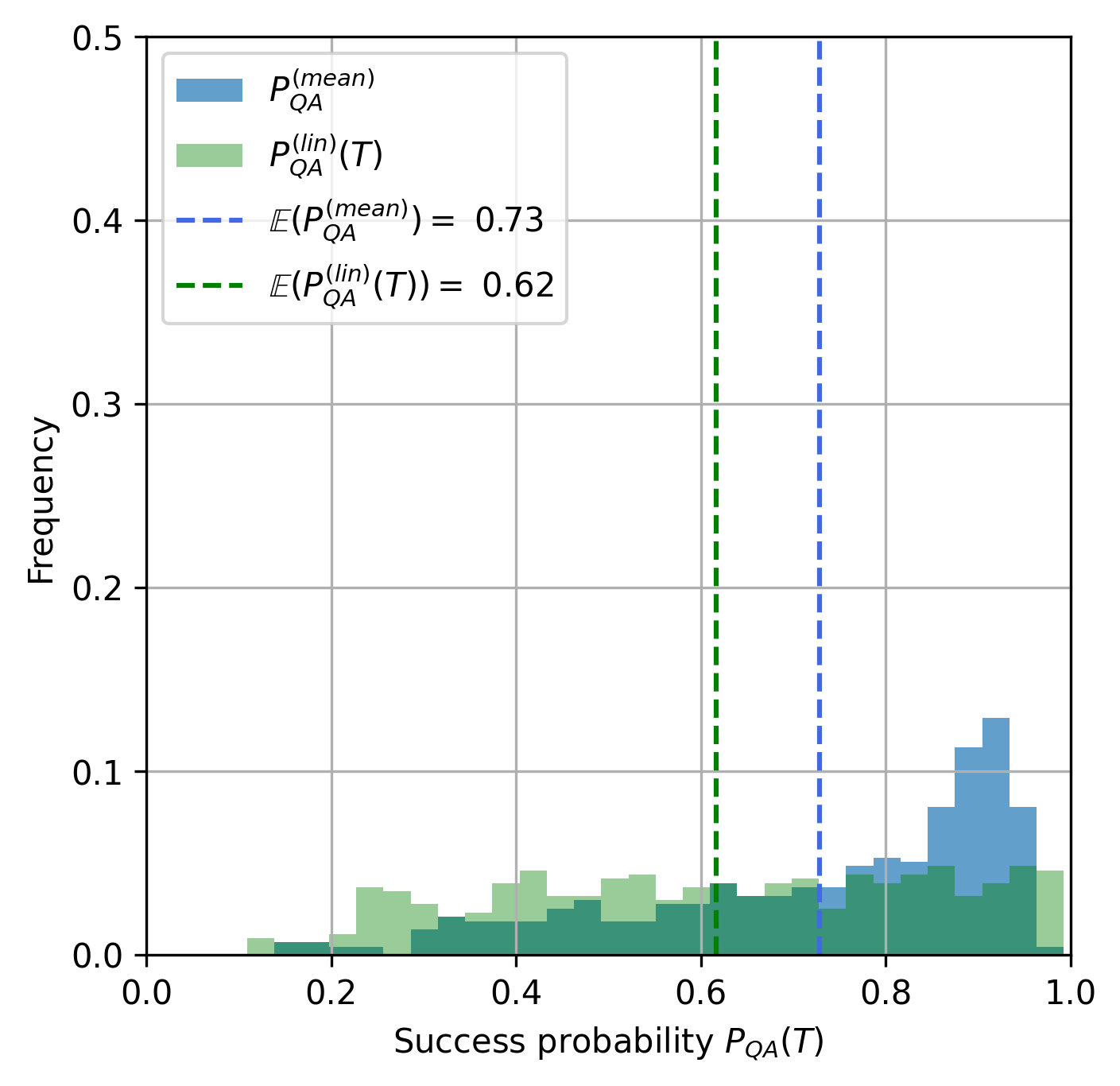}}\hspace{0.5cm}
    \subfloat[$N=8$ \label{subFig:histoMeanGapLinN8_snr15}]{%
    \includegraphics[scale=0.3]{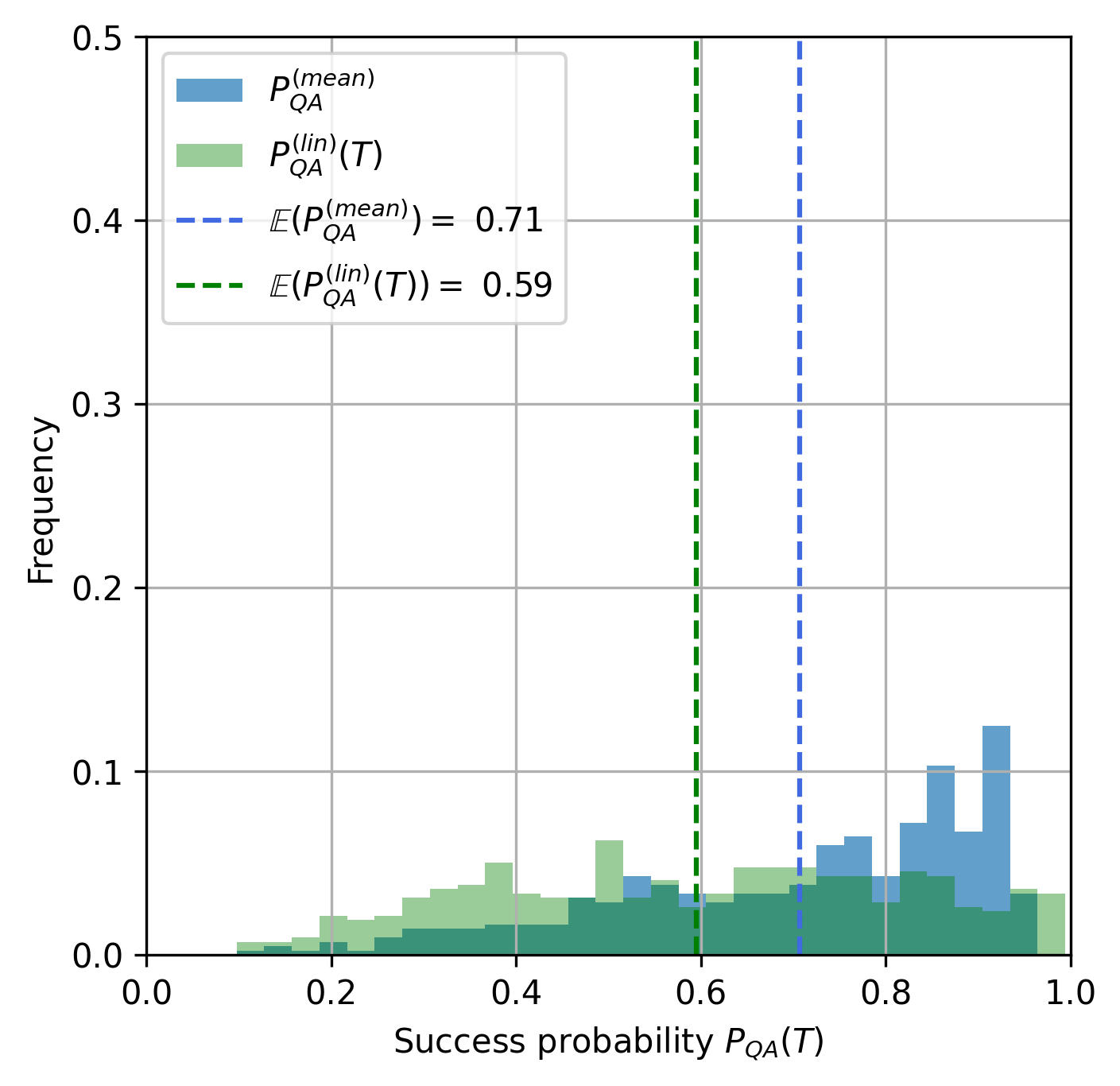}}\hspace{0.5cm}
    \subfloat[$N=9$ \label{subFig:histoMeanGapLinN9_snr15}]{%
    \includegraphics[scale=0.3]{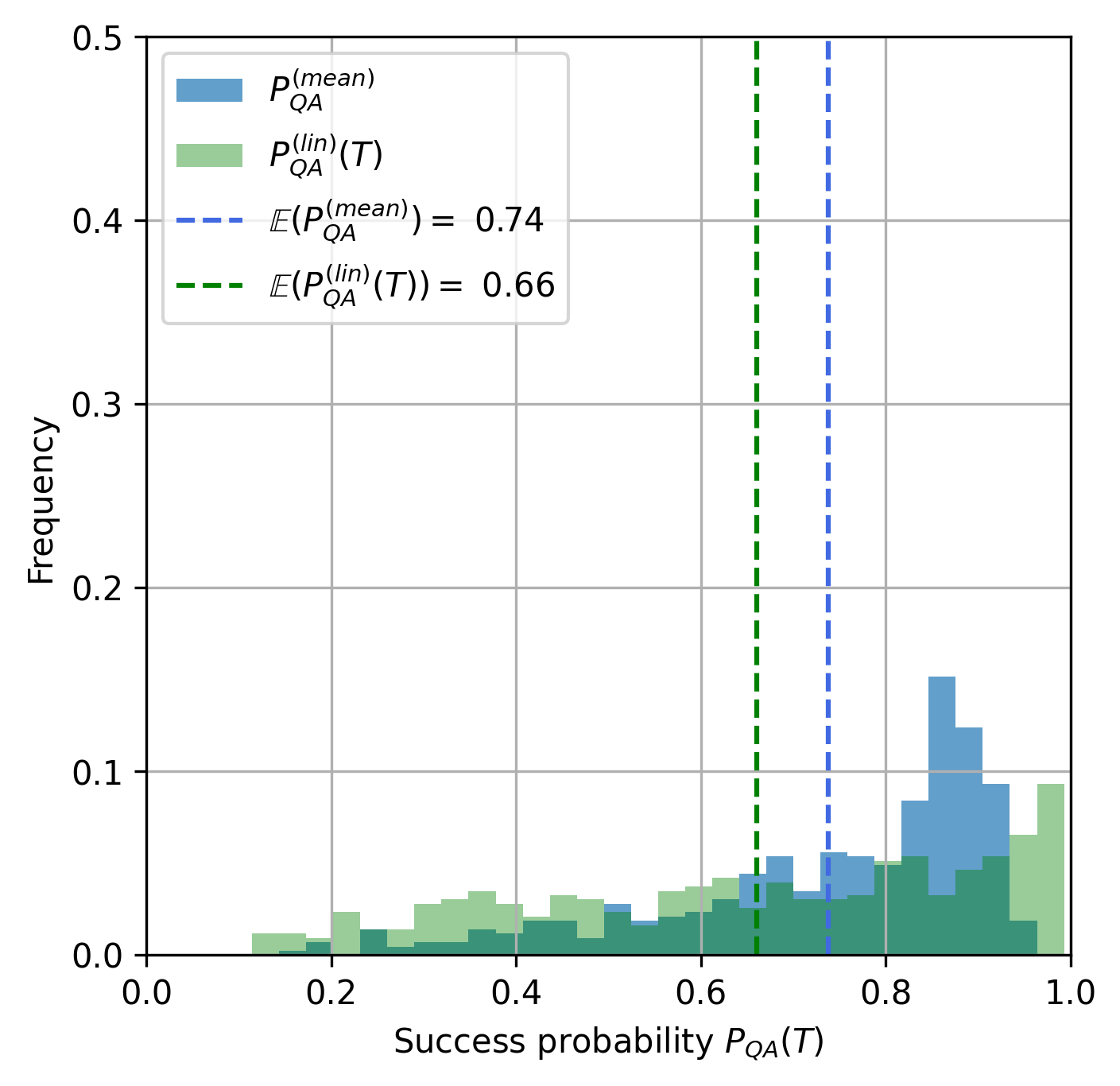}}
  \caption{Same distributions than Fig. \ref{fig:LinVsMeanNoChannel20db} but with $\text{SNR} = 15\text{dB}$. \hspace{30cm} $\,\,\,\,\,\,\,\,\,\,\,\,\,\,\,\,\,\,\,\,\,\,\,\,\,\,\,\,\,\,\,\,\,\,\,\,\,\,\,\,\,\,\,\,\,\,\,\,\,\,\,\,\,\,\,\,\,\,$} 
  \label{fig:LinVsMeanNoChannel15db}    
\end{figure*}

\subsection{Advantage over a linear control function}

Once the annealing time $T_{(\text{mean}),N}$ is known thanks to Eq. \ref{intMeanAnnealingT}, one might question the necessity to use the control function $u_{(\text{mean}),N}$ on the annealing period $[0, T_{(\text{mean}),N}]$. Indeed, the differential equation \ref{ODEu} to obtain the optimized control function associated to a given gap results from several approximations. Thus, one might be tempted to simply use a linear control function with an annealing period $T_{\text{lin},N} = T_{(\text{mean}),N}$ and expect the resulting success probability to be not too low even if the condition of Eq. \ref{conditionTlin} is not satisfied. To answer this question, we define the success probability of a QA process planned according to this strategy:
\begin{align}
    P_{\text{QA}}\left(T_{(\text{lin}),N} = T_{(\text{mean}),N} ; (\bm{y},\bm{w})\right) \equiv P_{\text{QA}}^{(\text{lin})}(\bm{y},\bm{w}).
\end{align}

Following the same approach than above, we reported on Fig. \ref{fig:LinVsMeanNoChannel20db} the distributions of $P_{\text{QA}}^{(\text{lin})}$ and $P_{\text{QA}}^{(\text{mean})}$ for a level of noise corresponding to $\text{SNR} = 20\text{dB}$. The expectation value $\mathbb{E}_{(\bm{y},\bm{w})}\left(P_{\text{QA}}^{(\text{mean})}\right)$ is higher than $\mathbb{E}_{(\bm{y},\bm{w})}\left(P_{\text{QA}}^{(\text{lin})}\right)$ which confirms the advantage of using $u_{\text{mean},N}$.

As the noise increases, Fig. \ref{fig:LinVsMeanNoChannel15db} shows that the gap between the two expectation values $\mathbb{E}_{(\bm{y},\bm{w})}\left(P_{\text{QA}}^{(\text{mean})}\right)$ and $\mathbb{E}_{(\bm{y},\bm{w})}\left(P_{\text{QA}}^{(\text{lin})}\right)$ is reduced. Nevertheless at $\text{SNR} = 15\text{dB}$, our control function is still expected to yield a success probability higher than the one that would been obtained with a linear control function

\section{Control function in the general case}
\label{sec:controlFctChannel}
In the previous section we proposed a generic method to evaluate an appropriate control function and demonstrated its validity on random problem instances with thermal noise at the BS. However in more realistic scenarios, the received signal can also be impacted by the channel coefficients. In this section, we take into account the channels between the users and the BS. From now on, the channel coefficients are normal variables to account for Rayleigh fading:
\begin{align}
    \bm{w} \sim \mathcal{N}(\bm{0},I_N).
\end{align}

\subsection{Behavior of the gap}

\begin{figure}
\centering
    \subfloat[Eigenvalues $\varepsilon_0$ and $\varepsilon_1$ \label{subFig:energyLevelsN8Channel}]{%
    \includegraphics[scale=0.35]{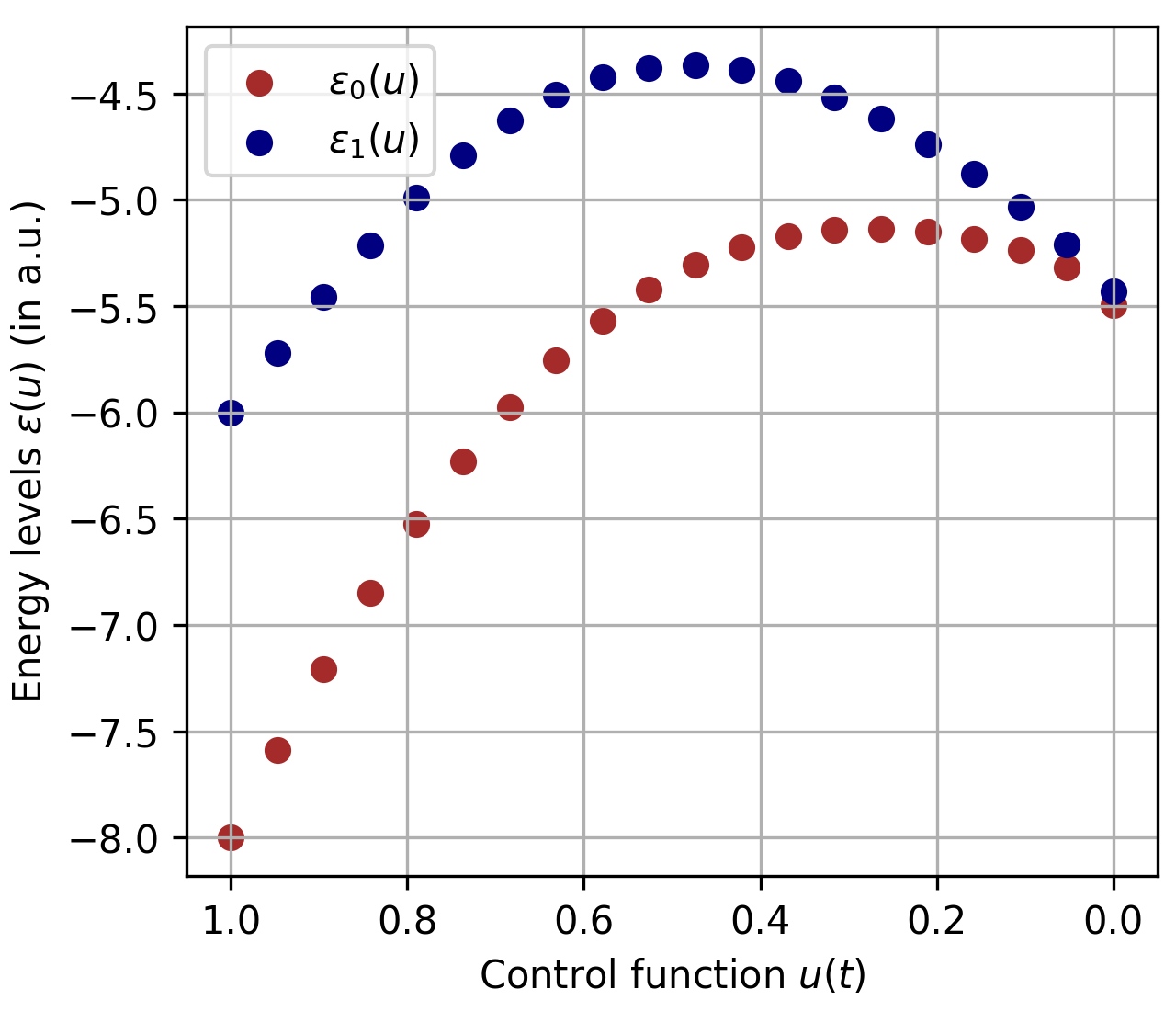}}\hspace{0.5cm}
    \subfloat[Evolution of $\Delta^2$ \label{subFig:gapN8Channel}]{%
    \includegraphics[scale=0.35]{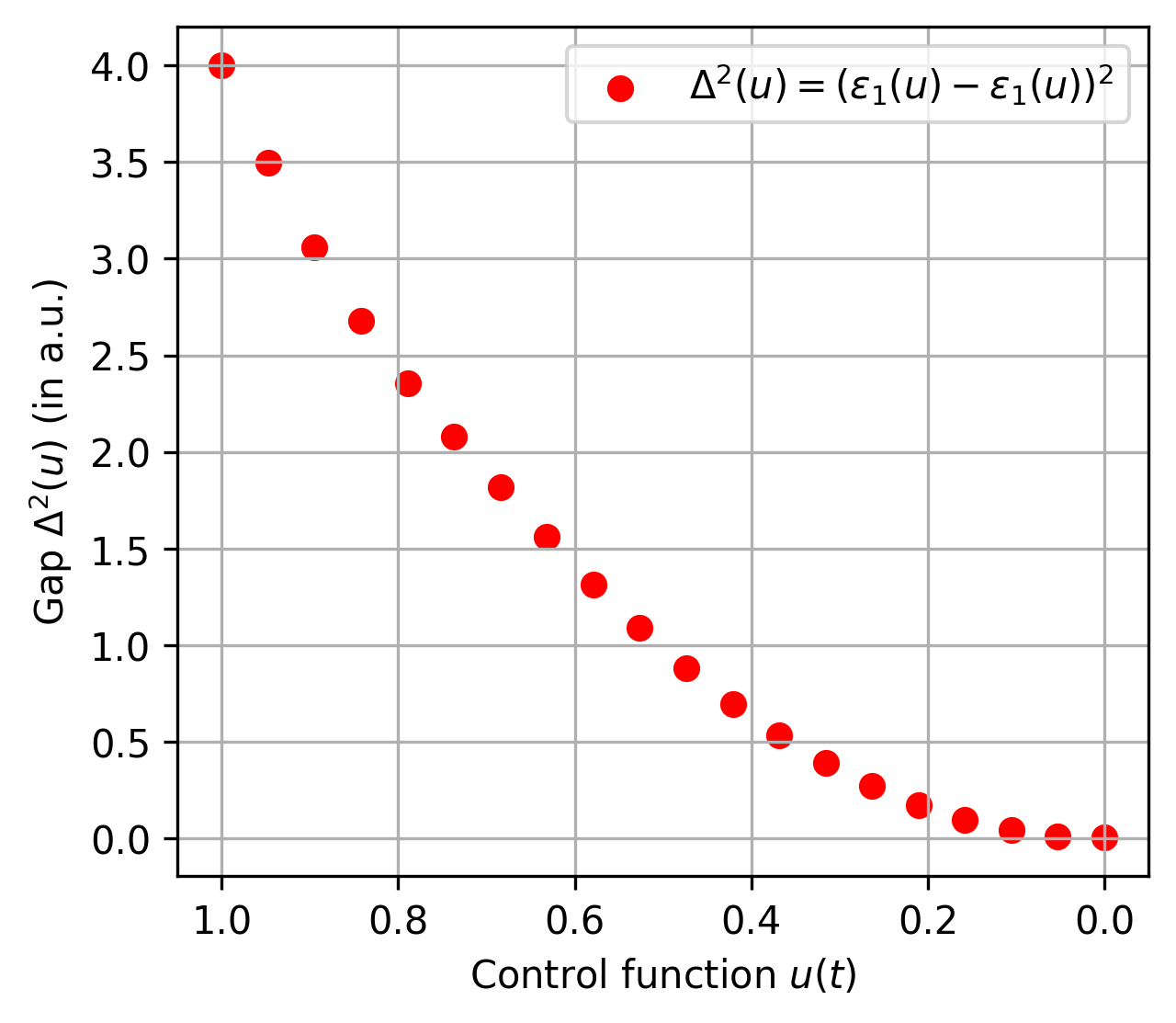}}\hspace{0.5cm}
  \caption{Two first eigenvalues of $\hat{H}(u)$ in a scenario with attenuation (a) and corresponding spectral gap against $u$ (b).} 
  \label{fig:gapChannel}    
\end{figure}

\begin{figure}
    \centering
    \includegraphics[scale=0.4]{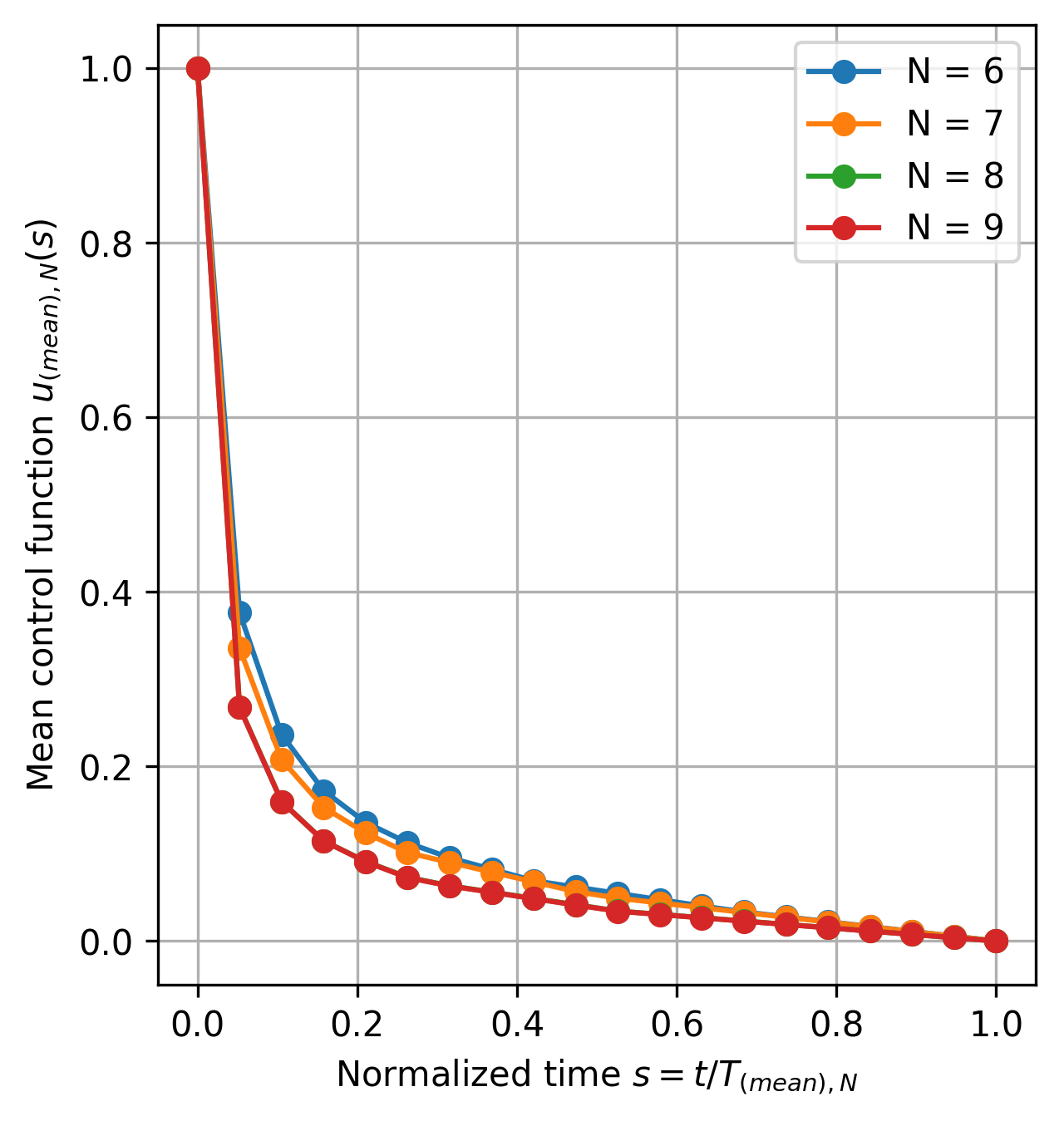}
    \caption{Shape of the mean control functions with fading coefficients in the network obtained for $N=6,\dots,9$ with respect to $s = t/T_{(\text{mean}),N}$}
    \label{fig:meanControlFctMutliNChannel}
\end{figure}

\begin{figure*}
\centering
    \subfloat[$N=6$ \label{subFig:histoMeanGapOptiChannel_N6_snr15}]{%
    \includegraphics[scale=0.3]{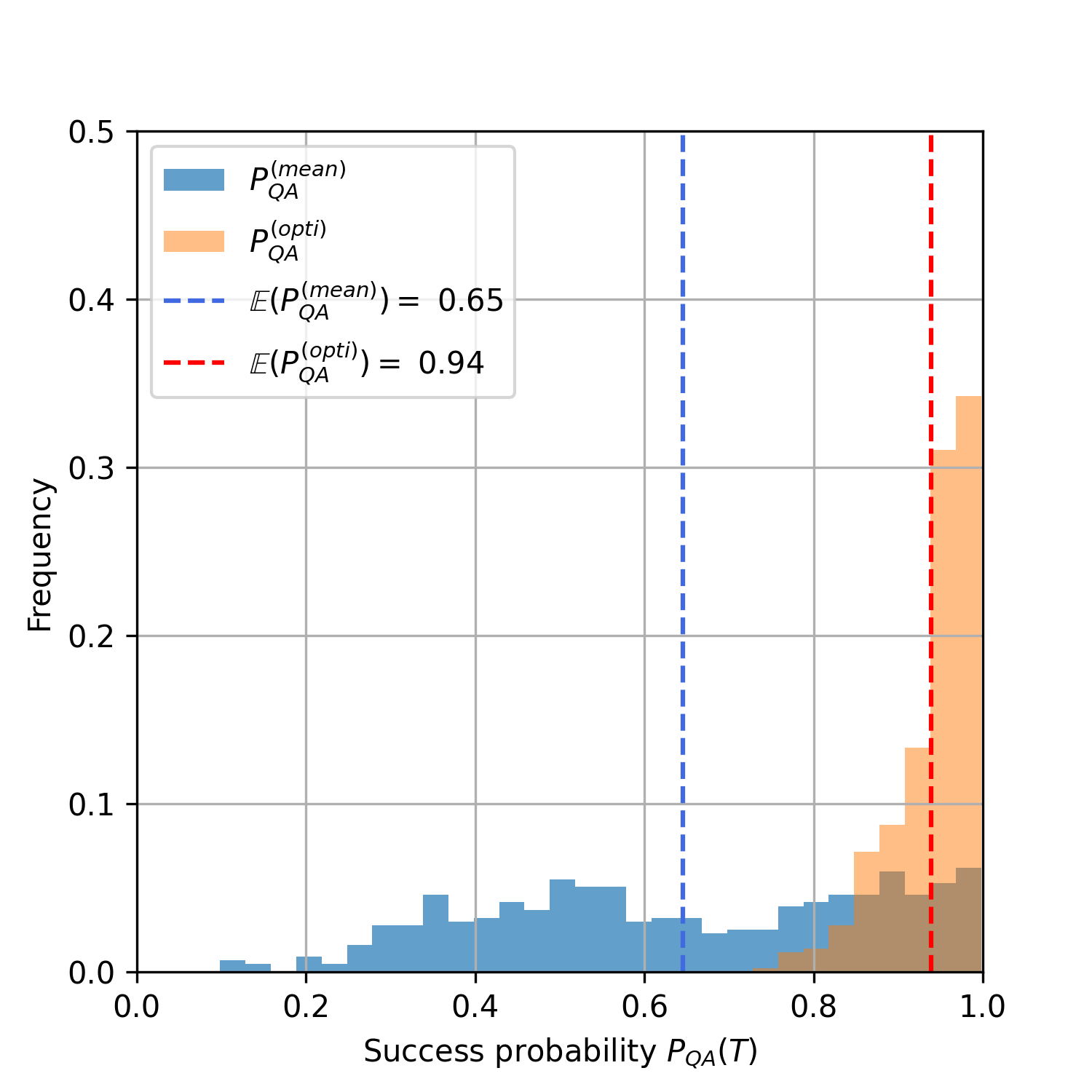}}\hspace{0.5cm}
    \subfloat[$N=7$ \label{subFig:histoMeanGapOptiChannel_N7_snr150}]{%
    \includegraphics[scale=0.3]{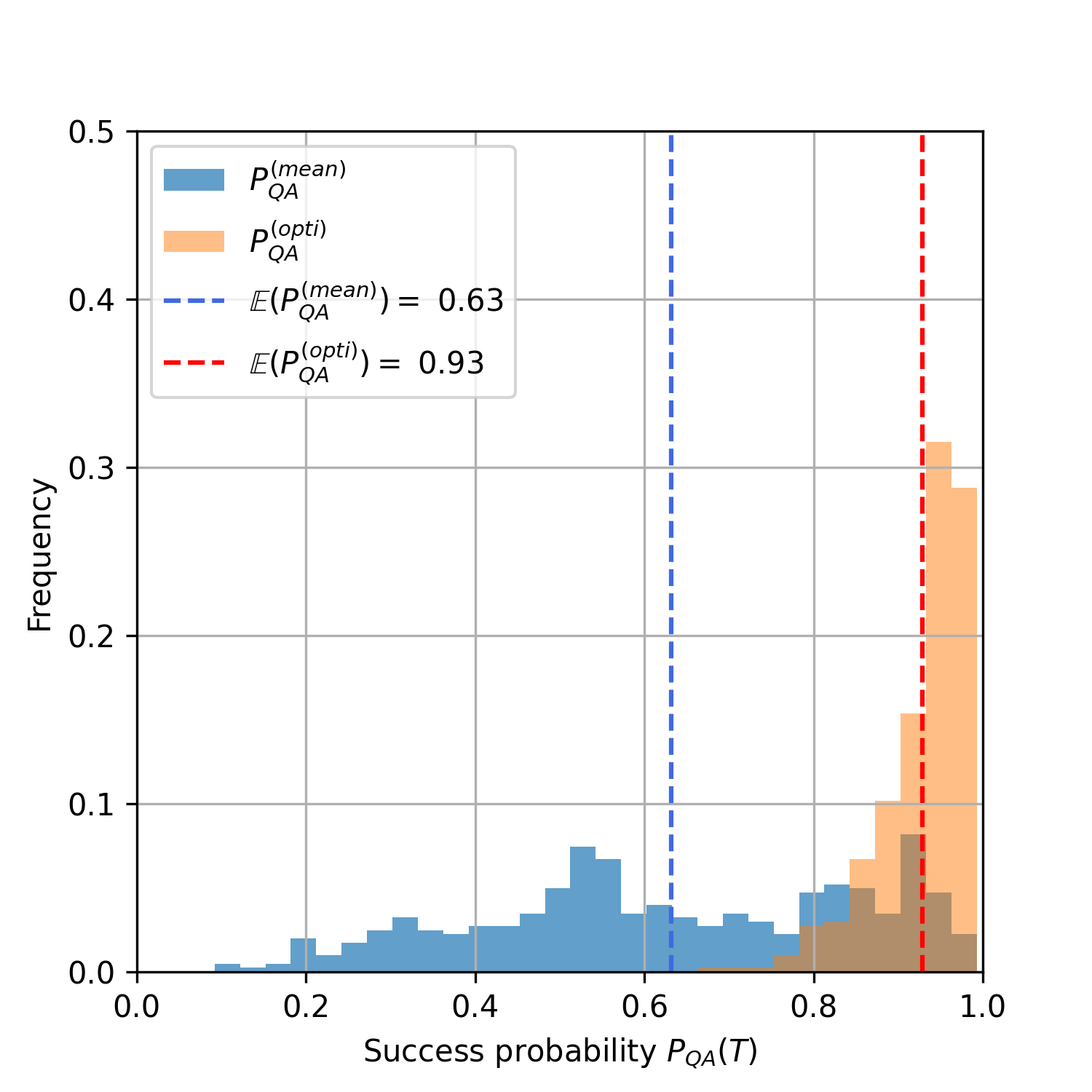}}\hspace{0.5cm}
    \subfloat[$N=8$ \label{subFig:histoMeanGapOptiChannel_N8_snr15}]{%
    \includegraphics[scale=0.3]{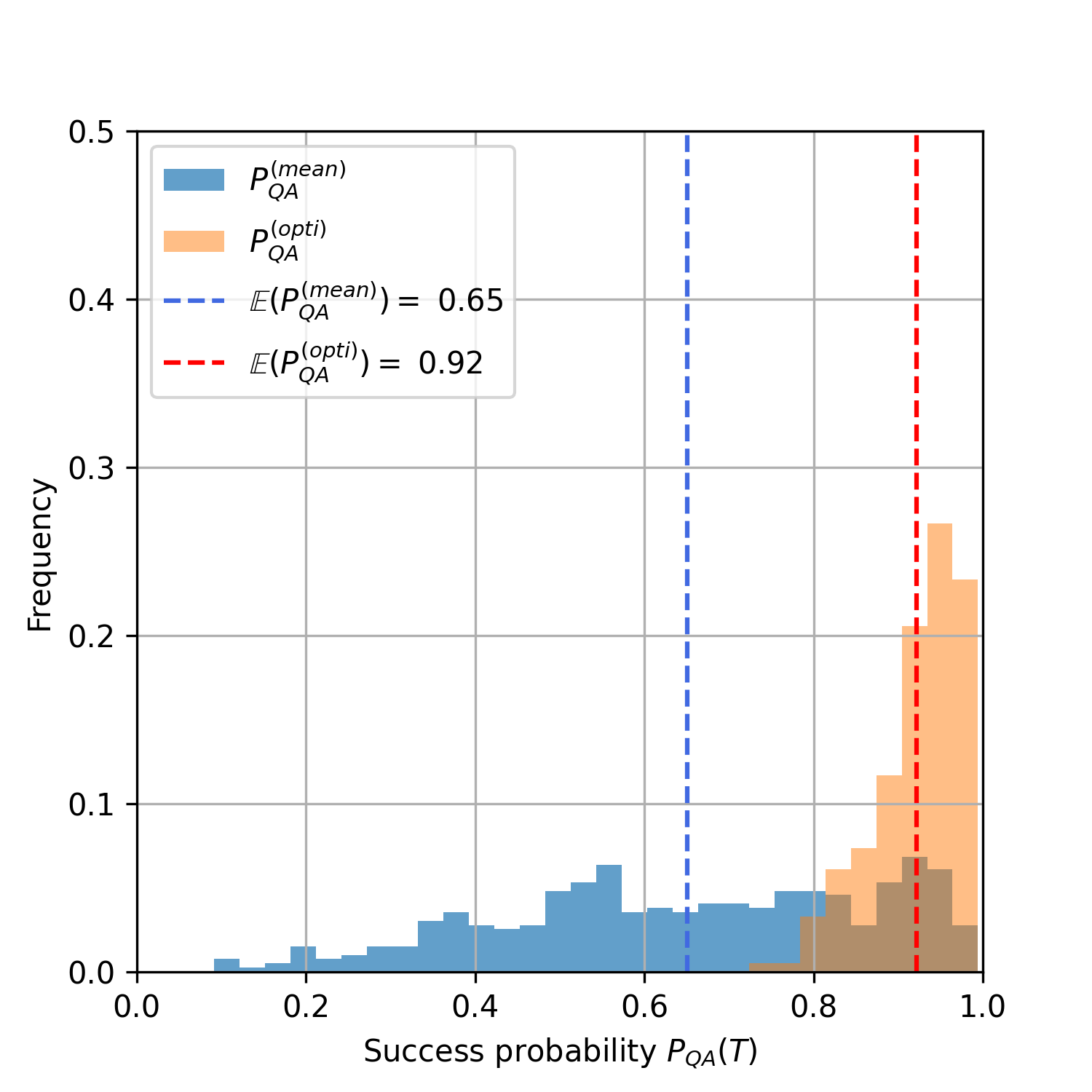}}\hspace{0.5cm}
    \subfloat[$N=9$ \label{subFig:histoMeanGapOptiChannel_N9_snr15}]{%
    \includegraphics[scale=0.3]{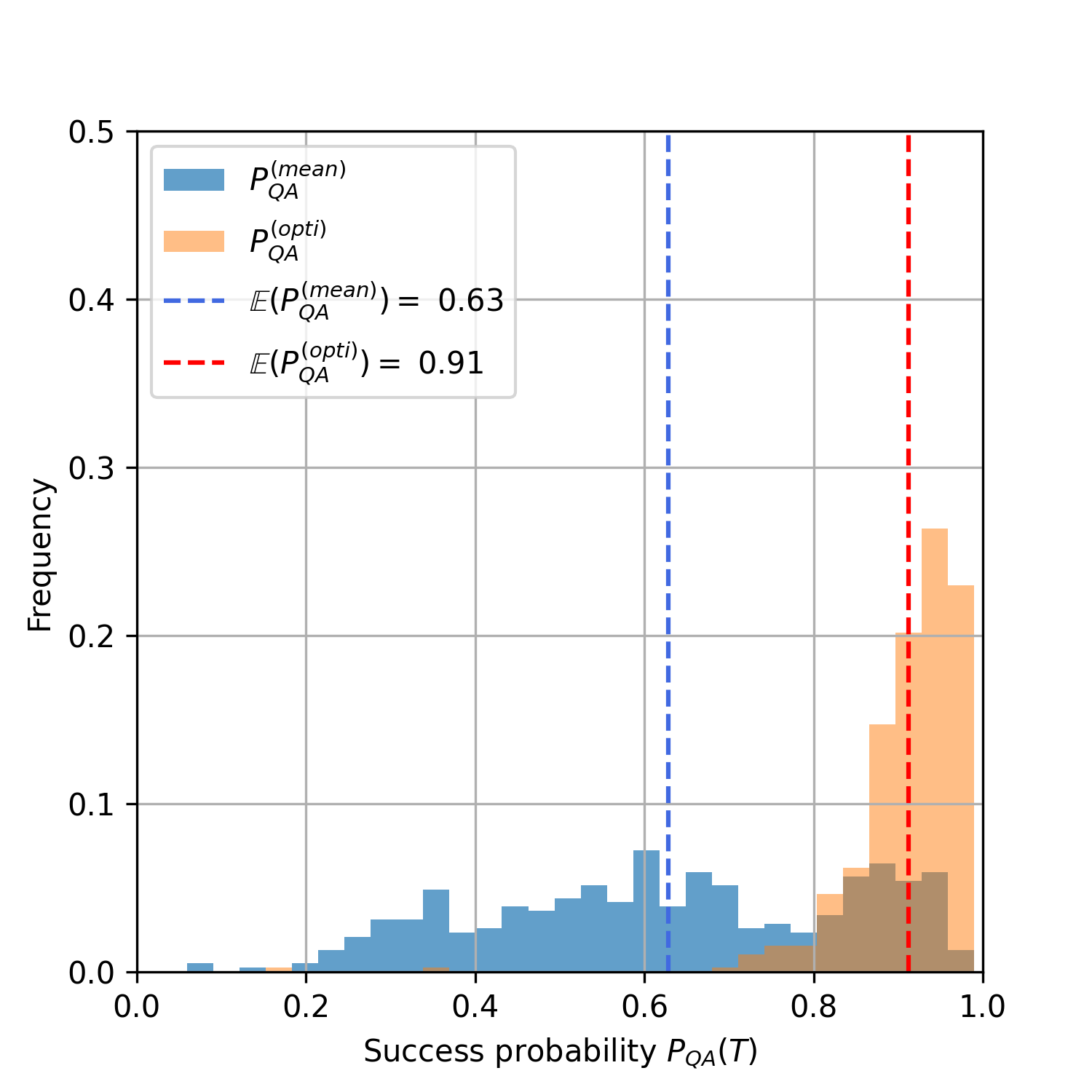}}
  \caption{Distributions of $P_{\text{QA}}^{(\text{mean})}$ (in blue) and $P_{\text{QA}}^{(\text{opti})}$ (in red) for several network sizes $N$ with channel imperfections taken into account. The noise level corresponds to $\text{SNR} = 15 \text{dB}$. We performed $N_{\text{samples}} = 500$ samples for each histogram.} 
  \label{fig:OptiVsMeanChannel15db}    
\end{figure*}

First of all, let us go back to our scenario of $N=8$ users with an initial activity pattern $\bm{b}^{(0)} = (1,0,0,0,0,0,0,0)$. We generate one random problem instance $(\bm{y},\bm{w})$ with no additional Gaussian noise by independently sampling the $N$ channel coefficients according to the standard normal distribution.

As done before, we reported on Fig. \ref{fig:gapChannel} the eigenvalues $\varepsilon_{0,1}$ of the global Hamiltonian $\hat{H}$ and the corresponding spectral gap. This time, the minimum of the gap is reached at the end of the annealing process when the control function vanishes.

This new behavior can be explained by the modifications induced in the spectrum of the Ising Hamiltonian $\hat{H}_P$. In the case where a user $i$ is strongly attenuated by a coefficient $w_i$ close from 0, the spin flip $\sigma_i \rightarrow -\sigma_i$ will not significantly modify the associated eigenenergy. Thus, taking into account the attenuation on the propagation paths between the users and the access point strongly reduces the gaps between the eigenvalues of the associated Ising Hamiltonian $\hat{H}_P$. It suggests that the mean gap should be estimated again in this new scenario.

\subsection{Mean control function with fading coefficients}

As done previously, we first neglect the additive Gaussian noise in order to evaluate the mean gap from several problem instances. From this quantity, we derive the associated mean control function $u_{(\text{mean}),N}$ defined on $[0,T_{(\text{mean}),N}]$. Since the channel coefficients $\bm{w}$ are now also randomly distributed, Eq. \ref{meanGapNoChannel} is generalized to:
\begin{align}
    \Delta^2_{(\text{mean}),N}(u) = \sum_{\bm{b}^{(0)} \in \{0,1\}^N} \int d\bm{w} f(\bm{b}^{(0)},\bm{w}) \Delta^2_{(\bm{y},\bm{w})}(u).
    \label{meanGapChannel}
\end{align}
Since $\bm{b}^{(0)}$ and $\bm{w}$ are independently distributed, the above probability density function is simply given by:
\begin{align}
    f(\bm{b}^{(0)},\bm{w}) = \frac{1}{2^N} \times \frac{1}{(2\pi)^{N/2}} \exp\left(-\frac{\lVert \bm{w} \rVert^2}{2}\right).
\end{align}
Contrarily to the scenario with perfect channels, the mean gap cannot be evaluated exactly through Eq. \ref{meanGapChannel}. Thus, we sample the probability distribution $f(\bm{b}^{(0)},\bm{w})$ with $N_{\text{samples, gap}} = 2000$ samples and evaluate the mean gap with the estimator:
\begin{align}
    \Delta^2_{(\text{mean}),N}(u) = \frac{1}{N_{\text{samples, gap}}} \sum_{(\bm{y},\bm{w}) \in \text{samples}} \Delta^2_{(\bm{y},\bm{w})}(u). 
    \label{meanGapChannel_estimator}
\end{align}
As done previously, we reported on Fig. \ref{fig:meanControlFctMutliNChannel} the shape of the control functions against the normalized time $s = t/T_{(\text{mean}),N}$ for $N = 6, \dots, 9$. These shapes indicate that the first derivative is monotonically decreasing over the annealing period, which is expected regarding the previous comments about the behavior of the gaps $\Delta^2_{(\bm{y},\bm{w})}$. Let us now check whether this approach still offers a good compromise in terms of success probability.

\begin{figure*}
\vspace{-0.5cm}
\centering
    \subfloat[$\lambda=1$ \label{subFig:histoMeanGapOptiChannel_N8_snr15_dilat1}]{%
    \includegraphics[scale=0.3]{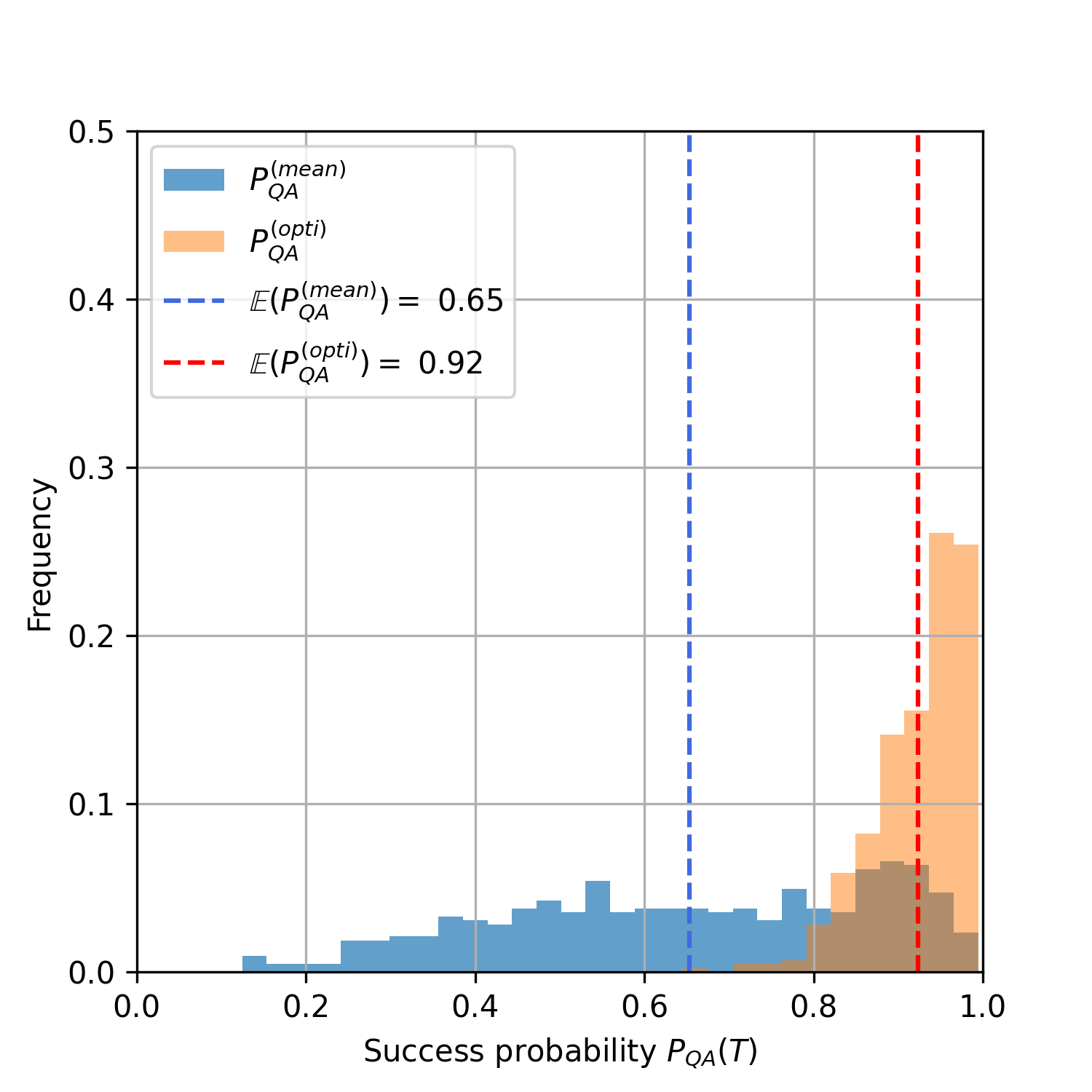}}\hspace{0.5cm}
    \subfloat[$\lambda=2$ \label{subFig:histoMeanGapOptiChannel_N8_snr15_dilat2}]{%
    \includegraphics[scale=0.3]{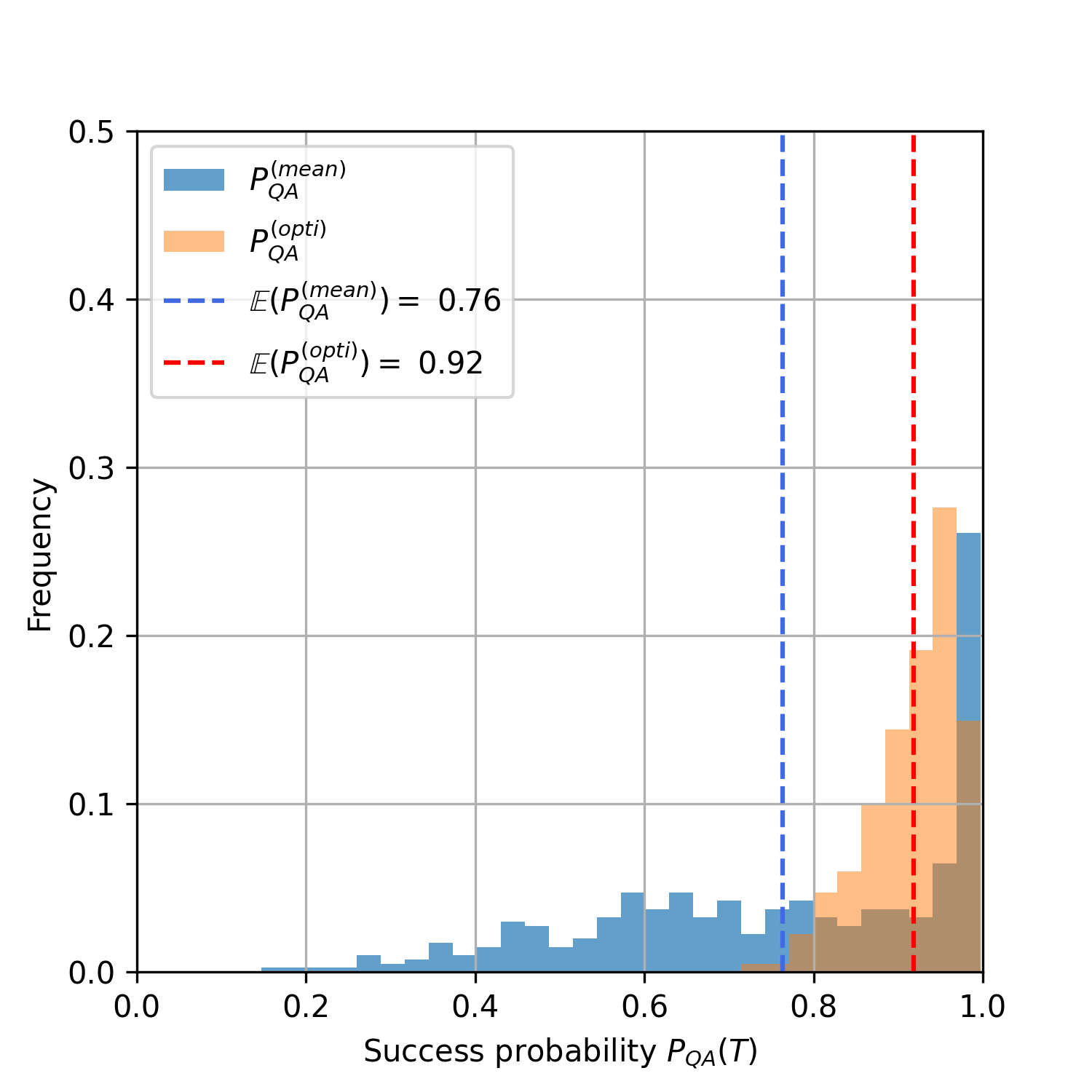}}\hspace{0.5cm}
    \subfloat[$\lambda=3$ \label{subFig:histoMeanGapOptiChannel_N8_snr15_dilat3}]{%
    \includegraphics[scale=0.3]{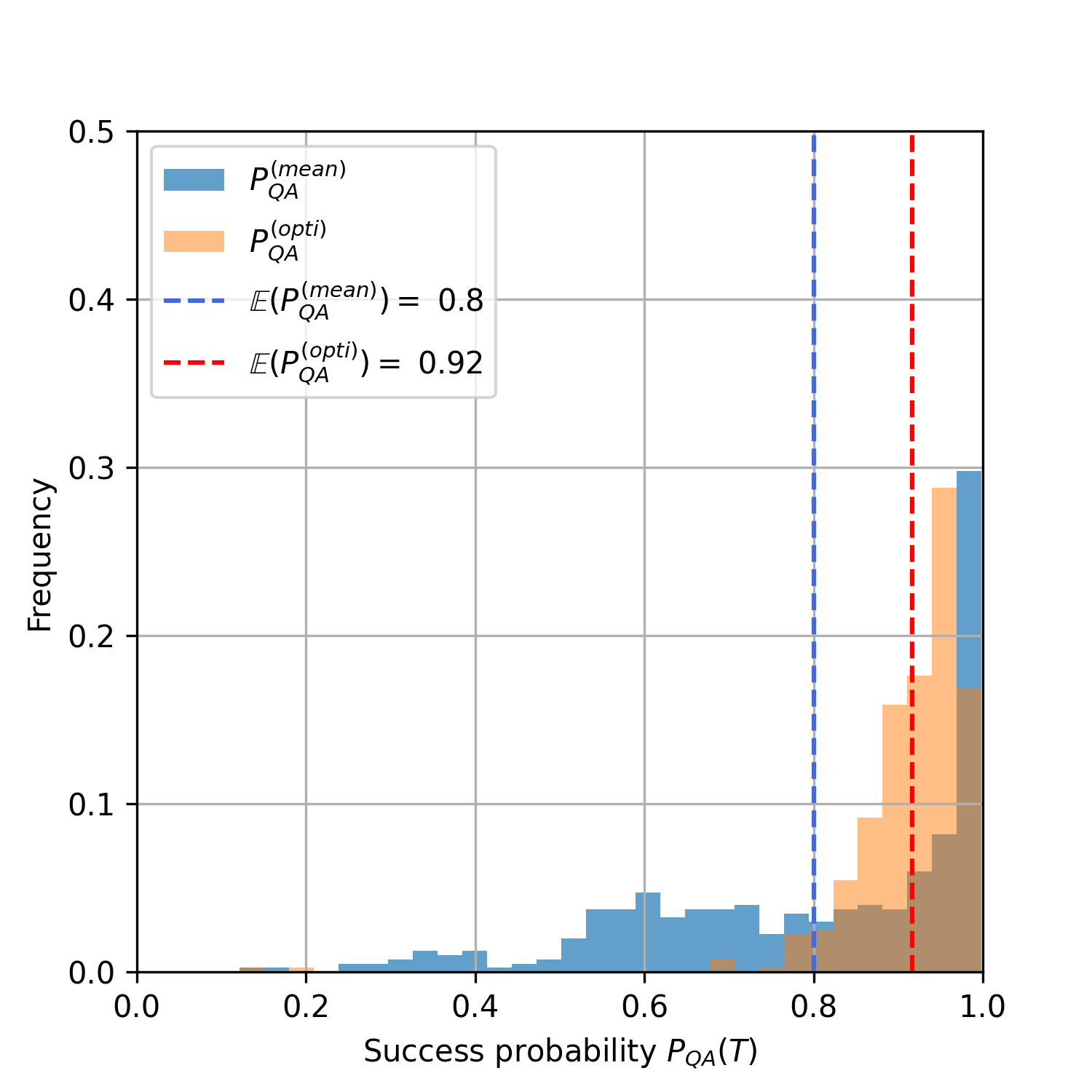}}\hspace{0.5cm}
    \subfloat[$\lambda=4$ \label{subFig:histoMeanGapOptiChannel_N8_snr15_dilat4}]{%
    \includegraphics[scale=0.3]{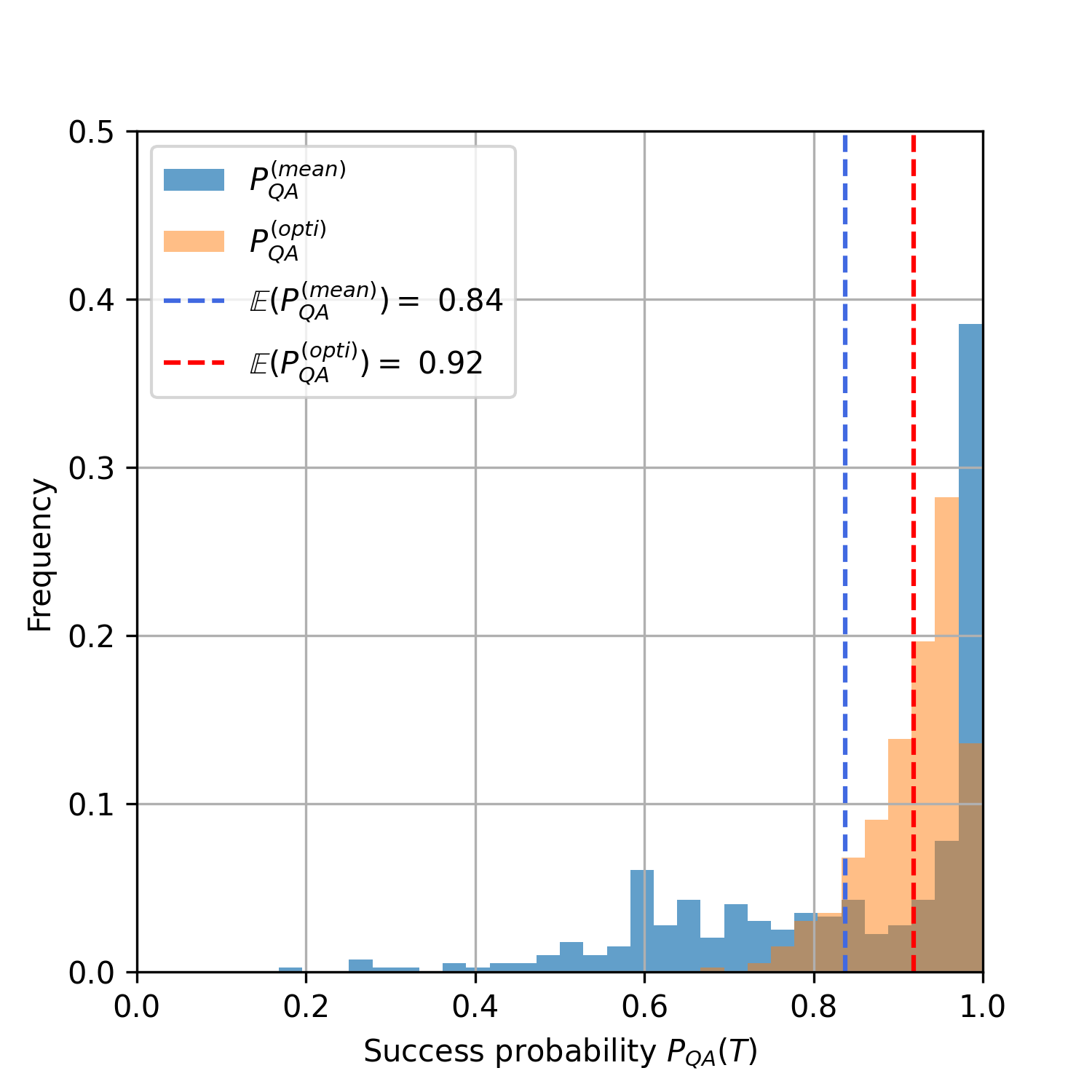}}
  \caption{Distributions $P_{\text{QA}}^{(\text{mean})}$ (in blue) and $P_{\text{QA}}^{(\text{opti})}$ (in red) for $N=8$ and $\text{SNR} = 15 \text{dB}$ with different dilatation factors $\lambda$. The annealing period $[0,T_{(\text{mean})}]$ is extended to $[0,\lambda T_{(\text{mean})}]$ and $[0,T_{(\bm{y},\bm{w})}]$ is left unchanged.}
  \label{fig:OptiVsMeanChannel15dBDilat}    
\end{figure*}

\begin{figure*}
\centering
    \subfloat[$\lambda=1$ \label{subFig:histoMeanGapLinChannel_N8_snr15_dilat1}]{%
    \includegraphics[scale=0.3]{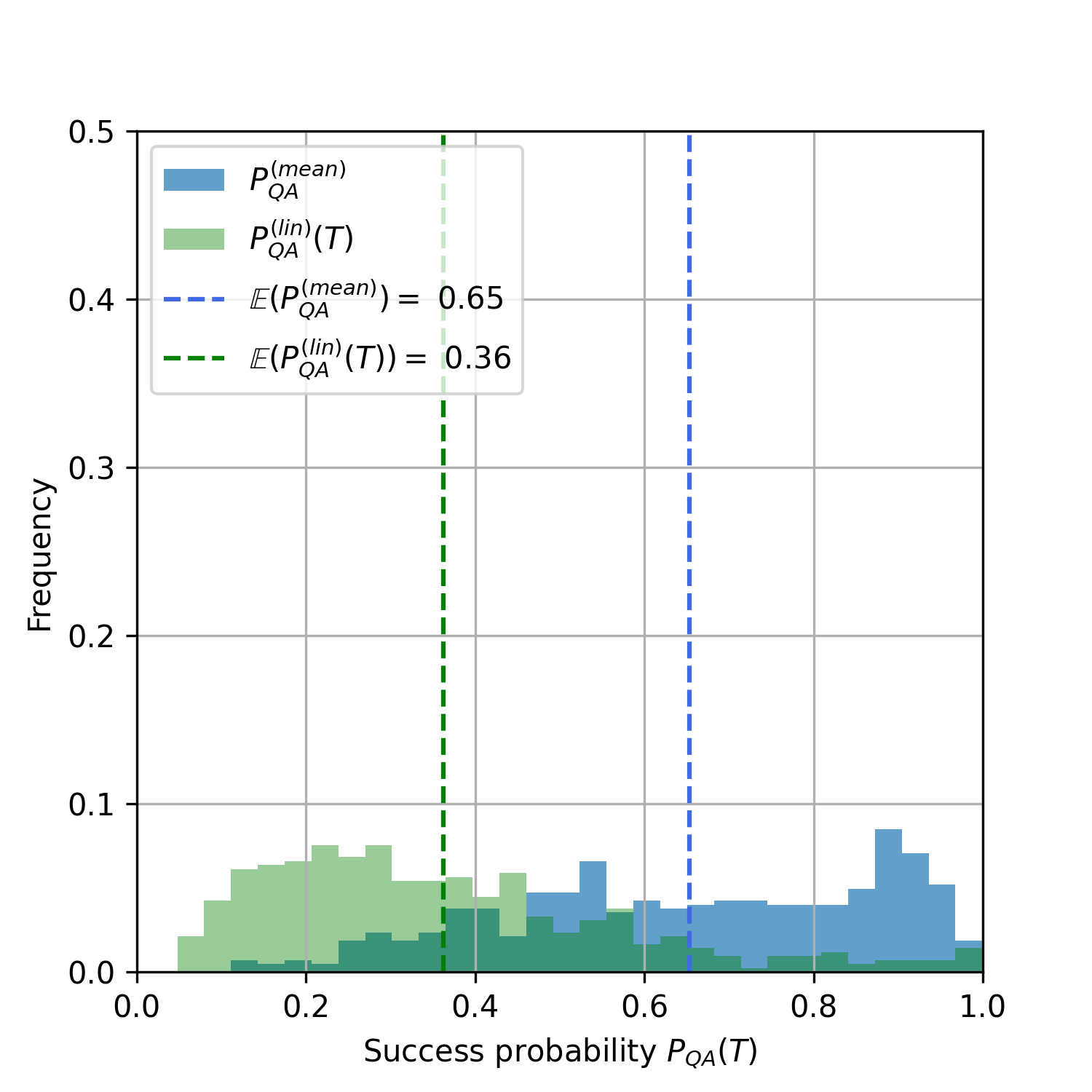}}\hspace{0.5cm}
    \subfloat[$\lambda=2$ \label{subFig:histoMeanGapLinChannel_N8_snr15_dilat2}]{%
    \includegraphics[scale=0.3]{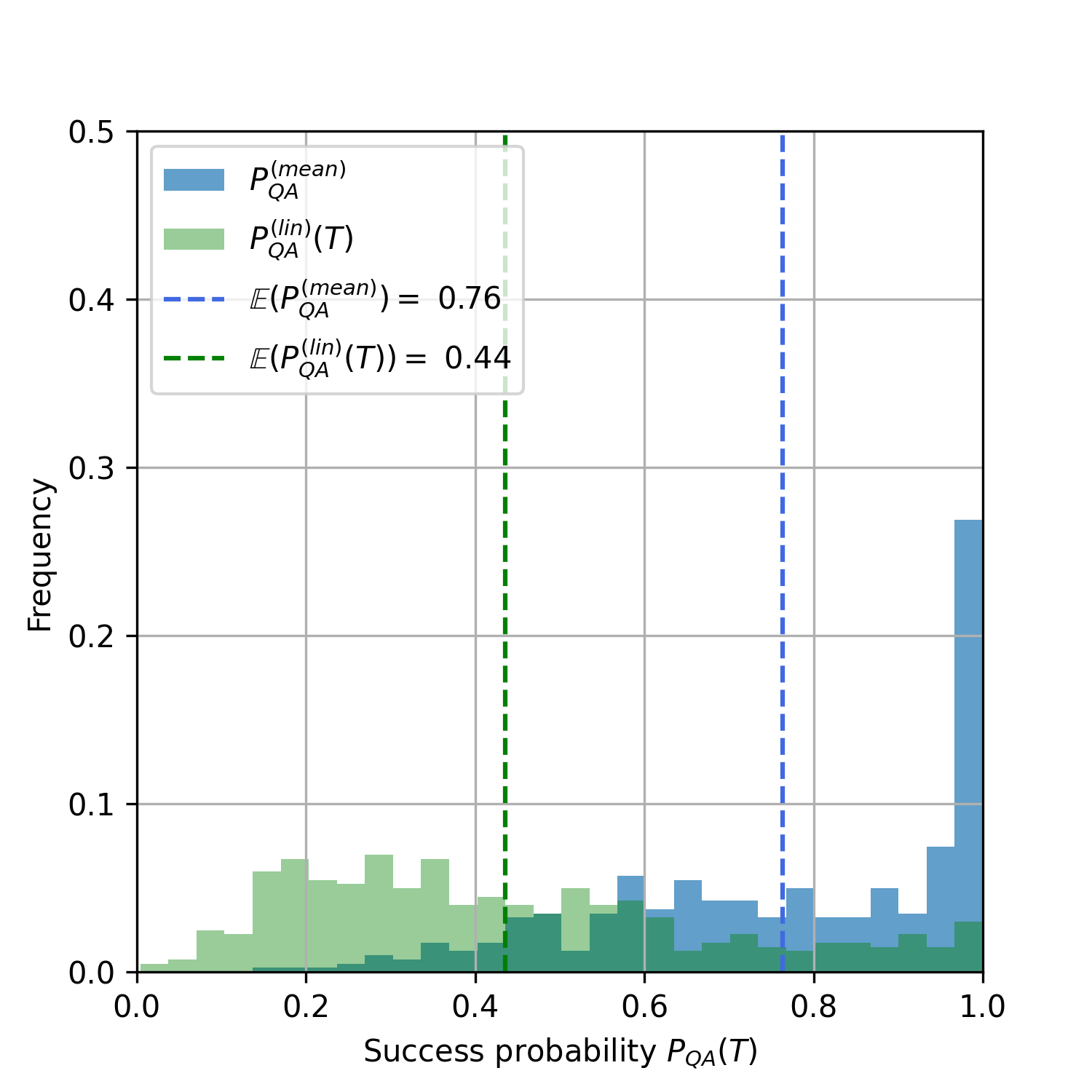}}\hspace{0.5cm}
    \subfloat[$\lambda=3$ \label{subFig:histoMeanGapLinChannel_N8_snr15_dilat3}]{%
    \includegraphics[scale=0.3]{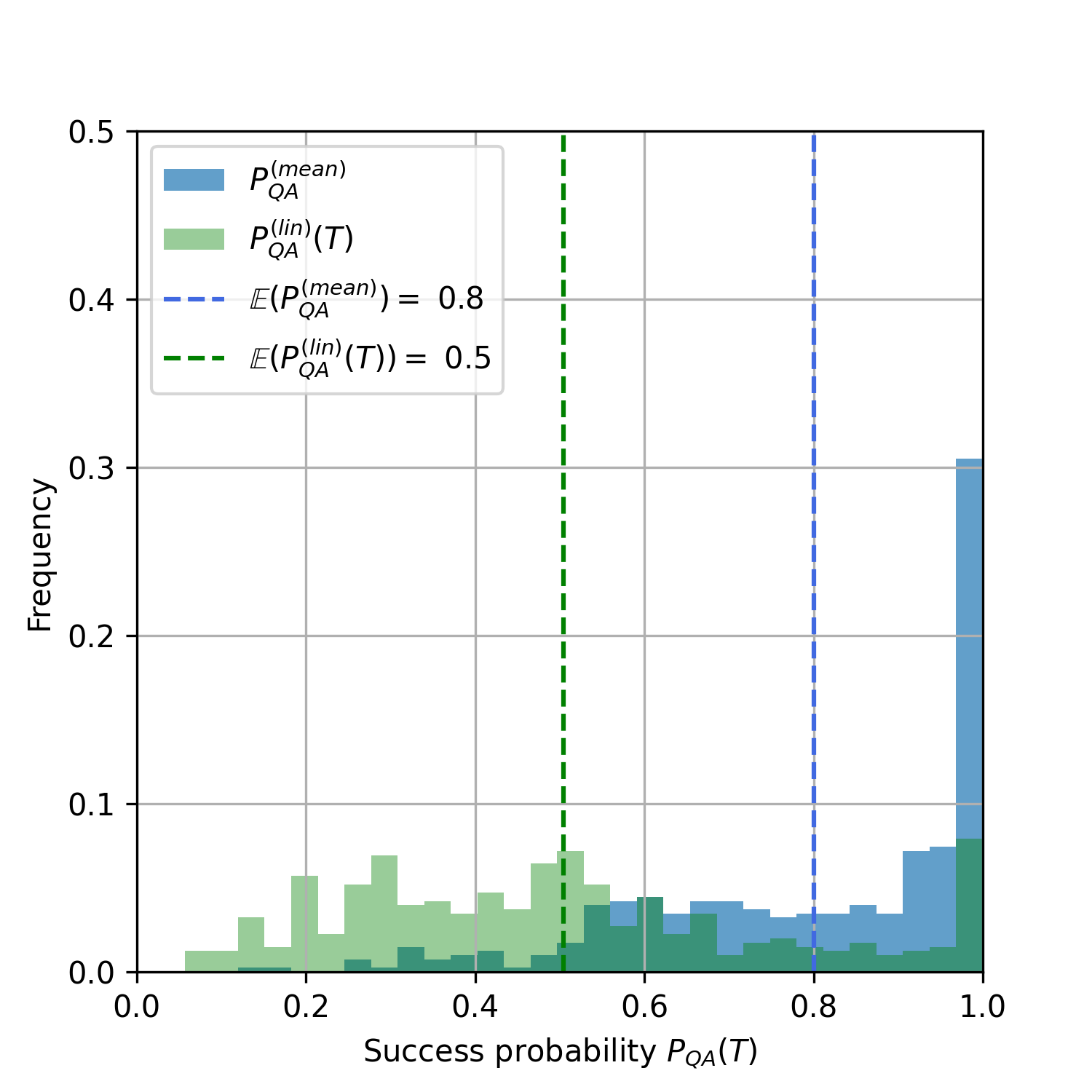}}\hspace{0.5cm}
    \subfloat[$\lambda=4$ \label{subFig:histoMeanGapLinChannel_N8_snr15_dilat4}]{%
    \includegraphics[scale=0.3]{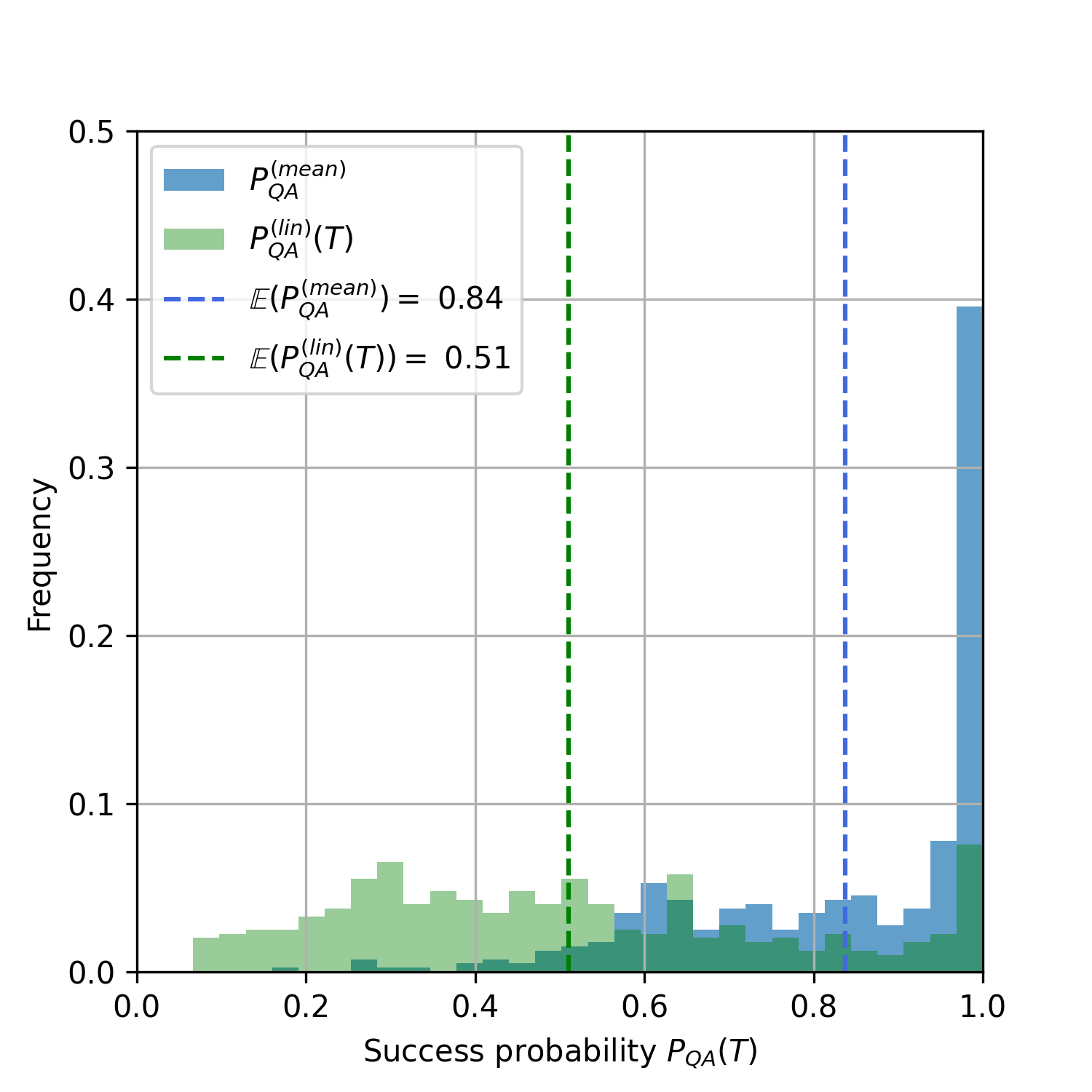}}
  \caption{Distributions $P_{\text{QA}}^{(\text{mean})}$ (in blue) and $P_{\text{QA}}^{(\text{lin})}$ (in green) for $N=8$ and $\text{SNR} = 15 \text{dB}$ with different dilatation factors $\lambda$.} 
  \label{fig:LinVsMeanChannel15dBDilat}    
\end{figure*}

\subsection{Mean success probability}

As we did with the perfect channel scenario, we would like to check that the mean control function still offers a good success probability when the noise is taken into account. We use again the metrics $P_{\text{QA}}^{(\text{mean})}$ and $P_{\text{QA}}^{(\text{opti})}$. The expression of their expectation values is modified according to the distribution of the instances $(\bm{y},\bm{w})$:
\begin{align}
\mathbb{E}_{(\bm{y},\bm{w})}\left(P_{\text{QA}}^{(\alpha)}\right) = \sum_{\bm{b}^{(0)} \in \{0,1\}^N} \int d\bm{n} d\bm{w} f(\bm{b}^{(0)}, \bm{n},\bm{w}) P_{\text{QA}}^{(\alpha)}(\bm{y},\bm{w}).
\end{align}
for $\alpha \in \{\text{mean},\text{opti}\}$. Using again the independence of the parameters used to generate a pair $(\bm{y},\bm{w})$, the above probability distribution is given by:
\begin{align}
\begin{split}
f(\bm{b}^{(0)},\bm{n},\bm{w}) & = \frac{1}{2^N} \times \frac{1}{(2\pi\xi)^{M/2}}  \exp\left(-\frac{\lVert \bm{n} \rVert^2}{2 \xi^2}\right) \\ \times & \frac{1}{(2\pi)^{N/2}} \exp\left(-\frac{\lVert \bm{w} \rVert^2}{2}\right). 
\end{split}
\end{align}

An estimation of the distributions of $P_{\text{QA}}^{(\text{mean})}$ and $P_{\text{QA}}^{(\text{opti})}$ at $\text{SNR} = 15\text{dB}$ is shown on the histograms of Fig. \ref{fig:OptiVsMeanChannel15db}. These results have been obtained by performing again $N_{\text{samples}} = 500$ samples of problem instances to evaluate $P_{\text{QA}}^{(\text{mean})}(\bm{y},\bm{w})$ and $P_{\text{QA}}^{(\text{opti})}(\bm{y},\bm{w})$. Unfortunately, the mean success probability $\mathbb{E}_{(\bm{y},\bm{w})}\left(P_{\text{QA}}^{(\text{mean})}\right)$ is around 0.65 while a well-suited control function for each problem instance still ensures a mean success probability $\mathbb{E}_{(\bm{y},\bm{w})}\left(P_{\text{QA}}^{(\text{opti})}\right)$ above 0.9. We must then investigate a way to improve the mean control function in this more complex AUD scenario.

\subsection{Time dilatation}

The reduction of the mean success probability $\mathbb{E}_{(\bm{y},\bm{w})}\left(P_{\text{QA}}^{(\text{mean})}\right)$ is likely due to the fact that $u_{(\text{mean}),N}$ decreases too fast at the end of the annealing process. Indeed, the example previously shown on Fig. \ref{fig:gapChannel} suggests that when the control function approaches 0, the gap closes which means that $u$ must decrease slowly enough to offer a good reliability. In order to force $u_{(\text{mean}),N}$ to decrease slower, we propose to apply a time dilatation factor $\lambda$ to its associated annealing period $[0, T_{(\text{mean}),N}]$. In practice, it amounts to use the time-dilated mean control function:
\begin{align}
    u_{(\text{mean}),\lambda,N}(t) = u_{(\text{mean}),N}\left(\frac{t}{\lambda}\right).
\end{align}
Consequently, the annealing period is extended to $[0, \lambda T_{(\text{mean}),N}]$. Of course, we aim at keeping $\lambda$ reasonably low to preserve the order of magnitude of the annealing time. 

Nevertheless, we show on Fig. \ref{fig:OptiVsMeanChannel15dBDilat} that time dilatation factors of $\lambda = 2,3,4$ already allow to boost the mean success probability $\mathbb{E}_{(\bm{y},\bm{w})}\left(P_{\text{QA}}^{(\text{mean})}\right)$. These results correspond to a network with $N=8$ users and a level of additive noise fixed to $\text{SNR} = 15\text{dB}$. For instance, picking $\lambda = 2$ allows to reach a mean success probability above 0.75 which is significantly better than the results obtained in the previous paragraph.

We also compared again our proposal to a linear control function parameterized over the same annealing period $[0,\lambda T_{(\text{mean}),N}]$. The results on Fig. \ref{fig:LinVsMeanChannel15dBDilat} confirm that $u_{(\text{mean}),\lambda,N}$ offers significantly a higher mean success probability $\mathbb{E}_{(\bm{y},\bm{w})}\left(P_{\text{QA}}^{(\text{mean})}\right)$ than a linear control function.

Thus, our strategy is also promising in this more complex AUD scenario. A time dilatation can be applied to the mean control function to improve the performances.

\section{Conclusion}

\subsection{Contributions}
In this work, we proposed a generic strategy to parameterize a QA process with the usual transverse Hamiltonian to compute the MAP estimator for AUD in a NOMA network. Despite the randomness of several parameters involved in the couplings and the local fields of the corresponding targeted Ising Hamiltonian, the evolution of the spectral gap of the global Hamiltonian against $u(t)$ follows a recurrent shape. We have shown that our mean control function ensures a reasonable expectation value for the success probability of QA after (if wanted) applying a time dilatation operation. This function has the advantage to be universal once the size of the network $N$ is fixed and permits to reach better performances than the simple linear control function.

\subsection{Scalability with $N$}
The work conducted in this paper is limited to small network sizes. For larger values of $N$, the computation of the quantum gaps like those involved in Eq. \ref{meanGapChannel_estimator} would be too long with the exact diagonalization method of the full Hamiltonian used in this work. A further improvement would be to use Quantum Monte Carlo methods as done by \cite{boixo_quantum_2014} to extract quantum gaps involved in the mean gap estimator in order to compute it for $N \geq 9$.

\subsection{Knowledge of the channel coefficients}
Regarding the AUD model, another improvement would be to relax some assumptions we made. In fact, the knowledge of the channel coefficients $\bm{w}$ at the access point is a strong assumption that does not hold true in practice. Some modern formulations of AUD assume that the \textit{statistics} of the channel vector $\bm{w}$ (which is a Gaussian law in this work) is known by the access point but not the \textit{realizations}. A further perspective to this work could be to adapt the signal processing model to such case.

\section*{Acknowledgment}

The authors would like to thank Lélio Chetot and Bruno Fedrici for helpful discussions.

This work was supported by the ANR under the France 2030 program, grant "NF-PERSEUS : ANR-22-PEFT-0004".

\appendices

\section{Choice of the sequences}
\label{seqChoice}

The identification sequences of the users must be chosen such that each activity pattern $\bm{b}^{(0)}$ yields a different signal $\bm{y}$ in the absence of imperfections. It means that in the case $\xi = 0$ and $\bm{w} = \bm{1}_N$, one requires:
\begin{align}
\begin{split}
    \left\{\bm{y}\left(\bm{b}^{(0)},\bm{n} = \bm{0}, \bm{w}=\bm{1}_N\right) \right.&= \left. \bm{y}\left(\bm{b}^{(1)},\bm{n} = \bm{0}, \bm{w}=\bm{1}_N\right)\right\} \\
    \Rightarrow & \left\{\bm{b}^{(0)} = \bm{b}^{(1)}\right\}.
\end{split}
\end{align}
One can show that this condition is ensured if and only if:
\begin{align}
    \left\{\exists \bm{\lambda} \in \{-1,0,1\}^N \,\, \text{s.t} \,\, \sum_{i=1}^N \lambda_i \bm{c}_i = 0\right\} \Rightarrow \left\{\bm{\lambda} = 0\right\}.
    \label{conditionScheme}
\end{align}
Throughout this work, we used coding schemes constructed by a random selection of $N$ codes of size $M$ followed by an exhaustive verification of the condition \ref{conditionScheme}. We give below the matrices $\bm{C} = [\bm{c}_1 , \dots, \bm{c}_N] \in \mathbb{C}^{M \times N}$ corresponding to the codes we used for $N = 6,\dots,9$.
\begin{align*}
        \bm{C}_6 &= \frac{1}{\sqrt{5}}\begin{pmatrix}
        -1 & -1 & 1 & 1 & -1 & -1\\
        1 & 1 & -1 & 1 & 1 & -1 \\
        -1 & -1 & 1 & 1 & 1 & 1 \\
        1 & -1 & 1 & 1 & -1 & -1 \\
        -1 & 1 & 1 & 1 & -1 & 1 
    \end{pmatrix}, 
\end{align*}
\begin{align*}
\begin{split}
    \bm{C}_7 &= \frac{1}{\sqrt{6}}\begin{pmatrix}
        1 & 1 & 1 & 1 & -1 & -1 & -1 \\
        1 & 1 & 1 & 1 & 1 & -1 & 1 \\
        1 & -1 & -1 & 1 & -1 & 1 & 1 \\
        -1 & 1 & 1 & 1 & 1 & -1 & -1 \\
        1 & 1 & 1 & -1 & -1 & 1 & -1 \\
        -1 & -1 & 1 & -1 & -1 & -1 & 1
    \end{pmatrix}, \\
    \bm{C}_8 &= \frac{1}{\sqrt{6}} \begin{pmatrix}
        1 & 1 & -1 & 1 & 1 & -1 & 1 & -1 \\
        1 & -1 & -1 & -1 & 1 & -1 & 1 & 1 \\
        -1 & 1 & 1 & -1 & 1 & -1 & -1 & 1 \\
        1 & 1 & -1 & -1 & 1 & 1 & -1 & -1 \\
        1 & 1 & -1 & 1 & 1 & 1 & -1 & 1 \\
        -1 & 1 & -1 & 1 & -1 & -1 & -1 & -1
    \end{pmatrix}, \\
    \bm{C}_9 &= \frac{1}{\sqrt{7}} \begin{pmatrix}
        -1 & -1 & -1 & -1 & -1 & -1 & -1 & -1 & -1 \\
        -1 & -1 & 1 & -1 & -1 & -1 & 1 & -1 & 1 \\
        -1 & -1 & -1 & 1 & -1 & 1 & -1 & -1 & 1 \\
        -1 & 1 & 1 & -1 & 1 & -1 & -1 & 1 & -1 \\
        -1 & -1 & 1 & 1 & 1 & -1 & 1 & -1 & 1 \\
        -1 & 1 & 1 & 1 & 1 & 1 & 1 & 1 & 1 \\
        -1 & -1 & 1 & -1 & -1 & 1 & 1 & 1 & 1 
    \end{pmatrix}.
\end{split}
\end{align*}

\section{QUBO - Ising mapping}
\label{quboIntoIsing}
One uses the change of variable $\sigma_i = 1-2 b_i$ in the expression of the objective function \ref{QUBOformML} without taking into account the irrelevant constant $||\bm{y}||^2$:
\begin{align}
\begin{split}
    \left\lVert\bm{y}-\bm{C}\cdot \right.&\left. \text{diag}(\bm{w}) \cdot \bm{b}\right\rVert^2 \sim -2 \sum_{i=1}^N (\bm{y} \cdot w_i \bm{c}_i) \left(\frac{1}{2}(1-\sigma_i)\right) \\
    & + \sum_{i,j=1}^N (w_i \bm{c}_i \cdot w_j \bm{c}_j )\left(\frac{1}{2}(1-\sigma_i)\right) \left(\frac{1}{2}(1-\sigma_j)\right).
\end{split}
\end{align}
We throw away the constant terms (ie. not depending on the $\sigma$'s) to obtain:
\begin{align}
\begin{split}
    \left\lVert\bm{y}-\bm{C}\cdot \right.&\left. \text{diag}(\bm{w}) \cdot \bm{b}\right\rVert^2 \sim \sum_{i=1}^N \sigma_i \left( w_i \bm{y}\cdot \bm{c}_i - \frac{1}{2} w_i \bm{c}_i \cdot \sum_j w_j \bm{c}_j\right) \\
    & + \frac{1}{4} \sum_{i,j=1}^N (w_i w_j \bm{c}_i \cdot \bm{c}_j) \sigma_i \sigma_j.
\end{split}
\end{align}
Using that $\sigma_i^2 = 1$, the diagonal contribution of the double sum is also an irrelevant constant. Since the sum is symmetric under $i \leftrightarrow j$, we can use $\sum_{i \ne j} \equiv 2 \sum_{i < j}$. Hence the expression of the couplings in the main text:
\begin{align}
    J_{ij} = - \frac{1}{2} (w_i w_j) (\bm{c}_i \cdot \bm{c}_j).
\end{align}
As for the local fields, one can directly identify:
\begin{align}
    h_i = -w_i \bm{c}_i \cdot \left(\bm{y} - \frac{1}{2} \sum_j w_j \bm{c}_j\right).
\end{align}
It matches the expression of the main text:
\begin{align}
    h_i = -w_i \bm{c}_i \cdot \tilde{\bm{y}},
\end{align}
with the signal $\bm{\tilde{y}}$ expressed as in Eq. \ref{yTildeDef}
\newpage
\nocite{*}
\AtNextBibliography{\small}
\printbibliography

\end{document}